\documentclass{article}%
\usepackage{amsmath}
\usepackage{amssymb}
\usepackage{amsfonts}
\usepackage{graphicx}%
\newtheorem{theorem}{Theorem}

\newtheorem{definition}[theorem]{Definition}

\newtheorem{remark}[theorem]{Remark}

\begin{document}

\author{Waldyr A. Rodrigues, Jr.\\\hspace{-0.1cm}Institute of Mathematics, Statistics e Scientific Computation\\IMECC-UNICAMP CP 6065\\13083-970 Campinas, SP\\Brazil\\e-mail:walrod@mpc.com.br or walrod@ime.unicamp.br}
\title{Algebraic and Dirac-Hestenes Spinors and Spinor Fields \thanks{published:
\textit{Journal of Mathematical Physics} \textbf{45}(7), 2908-2944 (2004).}}
\date{last revised: 04/06/2004 (final version)}
\maketitle

\begin{abstract}
Almost all presentations of Dirac theory in first or second quantization in
Physics (and Mathematics) textbooks make use of covariant Dirac spinor fields.
An exception is the presentation of that theory (first quantization) offered
originally by Hestenes and now used by many authors. There, a new concept of
spinor field (as a sum of non homogeneous even multivectors fields) is used.
However, a carefully analysis (detailed below) shows that the original
Hestenes definition cannot be correct since it conflicts with the meaning of
the Fierz identities. In this paper we start a program dedicated to the
examination of the mathematical and physical basis for a comprehensive
definition of the objects used by Hestenes. In order to do that we give a
\emph{preliminary} definition of algebraic spinor fields (\emph{ASF}) and
Dirac-Hestenes spinor fields (\emph{DHSF}) on Minkowski spacetime as some
equivalence classes of pairs $(\Xi_{u},\psi_{_{\Xi_{u}}})$, where $\Xi_{u}$ is
a spinorial frame field and $\psi_{_{\Xi_{u}}}$ is an appropriate sum of
multivectors fields (to be specified below). The necessity of our definitions
are shown by a carefull analysis of possible formulations of Dirac theory and
the meaning of the set of Fierz identities associated with the `bilinear
covariants' (on Minkowski spacetime) made with \emph{ASF} or \emph{DHSF.} We
believe that the present paper clarifies some misunderstandings (past and
recent) appearing on the literature of the subject. It will be followed by a
sequel paper where definitive definitions of \emph{ASF} and \emph{DHSF }are
given as appropriate sections of a vector bundle called the \textit{left}
spin-Clifford bundle. The bundle formulation is essential in order to be
possible to produce a coherent theory for the covariant derivatives of these
fields on arbitrary Riemann-Cartan spacetimes. The present paper contains also
Appendices (A-E) which exhibits a truly useful collection of results
concerning the theory of Clifford algebras (including many `tricks of the
trade') necessary for the intelligibility of the text.

\end{abstract}
\tableofcontents

\section{Introduction}

Physicists usually make first contact with Dirac spinors and Dirac spinor
fields when they study relativistic quantum theory. At that stage they are
supposed to have had contact with a good introduction to relativity theory and
know the importance of the Lorentz and Poincar\'{e} groups. So, they are told
that Dirac spinors are elements of a complex $4$-dimensional space
$\mathbb{C}^{4}$, which are the carrier space of a particular representation
of the Lorentz group. They are told that when you do a Lorentz transformations
Dirac spinors behave in a certain way, which is different from the way vectors
and tensors behave under the same transformation. Dirac matrices are
introduced \ as certain matrices on $\mathbb{C(}4)$ satisfying certain
anticommutation rules and it is said that they close a particular Clifford
algebra, known as Dirac algebra. The next step is to introduce Dirac wave
functions. These are mappings, $\Psi:\mathcal{M}\rightarrow\mathbb{C}^{4}$,
from Minkowski spacetime $\mathcal{M}$ (at that stage often introduced as an
affine space) to the space $\mathbb{C}^{4}$. The set of all these mappings is
trestricted by imposing to it the structure of a Hilbert space. After that,
Dirac equation, which is a first order partial differential equation is
introduced for $\Psi(x).$ Physics come into play by interpreting $\Psi(x)$ as
the quantum wave function of the electron. Problems with this theory are
discussed and it is pointed out that the difficulties can only be solved in
relativistic quantum theory, where the Dirac spinor field, gains a new status.
It is no more simply a mapping $\Psi:\mathcal{M}\rightarrow\mathbb{C}^{4}$,
but a more complicated object (it becomes an operator valued distribution in a
given Hilbert space \footnote{See, e.g., \cite{streater} for a correct
characterization of these objects.}) whose expectation values on certain one
particle states can be represented by objects like $\Psi$. From a pragmatic
point of view, only this knowledge this is more than satisfactory. However,
that approach, we believe, is not a satisfactory one to any scientist with an
enquiring mind, in particular to one with is worried with the foundations of
quantum theory. For such person the first question which certainly occurs is:
what is the geometrical meaning of the Dirac spinor wave function. From where
did this concept came from?

Pure mathematicians, which study the theory of Clifford algebras, e.g., using
Chevalley's classical books \cite{3,4} learn that spinors are elements of
certain minimal \textit{ideals\footnote{Do not worry if you did not know the
meaning of this concept. It is not a difficult one and is introduced in the
Appendix. B.}} in Clifford algebras. In particular Dirac spinors are the
elements of a minimal ideal in a particular Clifford algebra, the Dirac
algebra. Of, course, the relation of that approach (\textit{algebraic
spinors)}, with the one learned by physicists (covariant spinors) is known
(see, e.g., \cite{5,18,19}), but is not well known by the great majority of
physicists, even for many which specialize in general relativity and more
advances theories, like string and \textit{M-}theory.

Now, the fact is that the algebraic spinor concept\footnote{Algebraic spinor
fields on Minkowski spacetime will be studied in details in what follows, and
in \cite{50} where the concept is introduced using fiber bunde theory on
general Lorentzian manifolds.} (as it is the case of the covariant spinor
concept) fail to reveal the true geometrical meaning of spinor in general and
Dirac spinors in particular.

In 1966, Hestenes \cite{29} introduced a new definition of spinor field, that
he called later \textit{operator} spinor field. These objects, which in this
paper, will be called Dirac-Hestenes spinor fields have been introduced by
Hestenes as mappings $\psi:\mathcal{M}\rightarrow\mathbb{R}_{1,3}^{0}$, where
$\mathbb{R}_{1,3}^{0}$ is the even subalgebra of $\mathbb{R}_{1,3}$, a
particular Clifford algebra, technically known as the \textit{spacetime
}algebra.\footnote{$\mathbb{R}_{1,3}$ is not the original Dirac algebra, which
is the Clifford algebra $\mathbb{R}_{4,1}$ ), but is closed related to it,
indeed $\mathbb{R}_{1,3}$ is the even subalgebra of the Dirac algebra (see the
Appendix B for details).} Hestenes in a series of remarkable papers
\cite{28,29h1,29h2,29h3,29h4,24guhestenes} applied his new concept of spinor
to the study of Dirac theory. He introduced an equation, now known as the
Dirac-Hestenes equation, which does \textit{not} contains (explicitly)
imaginary numbers and obtained a very clever interpretation of that theory
through the study of the geometrical meaning of the so called bilinear
covariants, which are the observables of the theory. He further developed an
interpretation of quantum theory from his formalism \cite{29h6,29h7}, that he
called the zitterbewegung interpretation. Also, he showed how his approach it
suggests a geometrical link between electromagnetism and the weak
interactions, different from the original one of the standard model
\cite{29h5}

Hestenes papers and his book with Sobczyk \cite{30} have been the inspiration
for a series of international conferences on \ `\textit{Clifford Algebras and
their Applications in Mathematical Physics}' which in 2002 have had its sixth
edition. A consultation of the table of contents of the last two conferences
\cite{ablamo1,rayan, ablamo2} certainly will show that Clifford algebras and
their applications generated a wider interest among many physicists,
mathematicians and \ even in engineering and computer sciences. Physicists
used\footnote{In what follows we quote some of the principal papers that we
have had opportunity to study. We apologise, any author that thinks that his
work is a worth one concerning the subject and is ot quoted in the present
article.} Clifford algebras concepts and Hestenes methods, in many different
applications. As some examples, we quote some developments in relativistic
quantum theory, as, e.g.,
\cite{challinor2,challinor3,daviau,72,deleo1,deleo2,deleo3,deleo4,deleo5,doran3,doran5,gull,furuta}%
. The papers by De Leo and collaborators, exhibit a close relationship between
Hestenes methods and quaternionic quantum mechanics, as developed, e.g., by
Adler \cite{adler}, a subject that is finding a renewed interest. Also,
Clifford algebra methods have been used
\cite{lasenby4,44pavrod,48WR1,48WR2,49WR1,49WR2,55a,vaz1,vaz2,vaz3} to give an
intuitive and geometrical clear picture of the dynamics of superparticles
\cite{6berezian}. Also, that papers clarify the meaning of Grassmann variables
and their calculus \cite{6beremari}. The relation with the zitterbewegung
model of Barut and collaborators \cite{2barut1,2barut2,2barut3} appear in a
novel and less speculative way. Even more, in \cite{49WR1} it is shown that
the concept of Dirac-Hestenes spinor field is closely related to the concepts
of superfields as introduced by Witten \cite{58}. Clifford algebras methods
have also been used in disclosing a surprising connection between the Dirac
and Maxwell and Seiberg-Witten \cite{57} equations, as studied, e.g., in
\cite{52,55,vaz3}, which suggest several physical developments. Applications
of Clifford algebras methods in general relativity appeared also, e.g., in
\cite{challinor1,29h8,doran1,doran2,doran4,doran5,fernandez,lasenby5,lasenby6,lasenby7,matteuci,oliveira,roqui}%
, and suggest new ways for looking to the gravitational field. Clifford
algebras methods, have been applied successfully also in quantum field theory,
as , e.g., in \cite{fauser,pavsic3} and more recently in string and
\textit{p}-brane theories, with noticeable results (\cite{castro1}%
-\cite{castro10},\cite{pavsic1,pavsic2}) which are worth to be more carefully investigated.

Of course, Clifford algebras and Dirac operators are a standard topics of
research in Mathematics (see, e.g., \cite{7}), but we must say that Hestenes
ideas have been an inspiring idea for mathematicians also. In particular, the
concept of Clifford valued functions with domain in a manifold (the operator
spinor fields are particular functions of this type) developed in a new,
beautiful and powerful branch of mathematics \cite{delanghe}. Hestenes ideas,
as we said, have found also their use in engineering and computer sciences, as
in the study of neural circuits \cite{29h9,29h10} and robotics and perception
action systems
\cite{bayco1,bayco2,lasenby1,lasenby2,colombo,dorst,lasenby3,41MOSNA3,sommer}.

Having making all this propaganda, which we hope have awaked the reader
interest in studying Clifford algebras, we must remark, that (as often happen
for every pioneer work) the concept of Dirac-Hestenes spinor field, as
originally introduced by Hestenes, and used by many other researchers, is
\textit{not} a concept free of criticisms and objections from the mathematical
point of view.

\ However, it is an important concept and one of the objectives of this paper
and also of \cite{50} is to give a presentation of the subject free of all
previous criticisms, which are discussed in the next sections. The reader, may
ask himself if the enterprising for learning the theory presented below is
worth his time. We think that the answer is yes, be him a physicist or a
mathematician. To encourage physicists, which may eventually become interested
in the subject after he read the above propaganda, we say that the
mathematical tools used, even if they may look complex at first sight, are
indeed nothing more than easy additions to the contents of a linear algebra
course. The main reward to someone that study what follows is that he will
start seeing some subjects that he thought were well known, under a new and
(we believe) illuminating point of view. This hopefully may help anyone who is
in the searching for new physical theories. For mathematicians, we say that
the point of view developed here is somewhat new in relation to the original
Chevalley's one and we believe, it is more satisfactory. In particular, the
present paper serves as a preliminary step towards a rigorous theory of
algebraic and Dirac-Hestenes spinor fields as sections of some well defined
fiber bundles, and the theory of the covariant derivatives of these fields.
Having said all that, what is the present paper about?

We give definitions of algebraic spinor fields (\emph{ASF}) and Dirac-Hestenes
spinor fields (\emph{DHSF}) living on Minkowski spacetime\footnote{Minkowski
spacetime is parallelizable and as such admits a spin structure. In general, a
spin structure does not exixt for an arbitrary manifold equipped with a metric
of signature $(p,q).$The conditions for existence of a spin structure in a
general manifold is discussed in \cite{31,42,44}, For the case of Lorentzian
manifolds, see \cite{20G}.} and show how Dirac theory can be formulated in
terms of these objects. We start our presentation in section 2 by studying a
not well known subject, namely, the geometrical equivalence of representation
modules of simple Clifford algebras $\mathcal{C}\ell(V,\mathbf{g)}$. This
concept, together with the concept of \textit{spinorial frames} play a crucial
role in our definition of algebraic spinors (\emph{AS) }and of \emph{ASF}.
Once we grasp the definition of \emph{AS} and particularly of Dirac \emph{AS
}we define Dirac-Hestenes spinors (\emph{DHS}) in section 4. Whereas \emph{AS
}may be associated to any real vector space of arbitrary dimension $n=p+q$
equipped with a non degenerated metric of arbitrary signature $(p,q)$, this is
not the case for \emph{DHS}\footnote{ASF can be defined on more general
manifolds called spin manifolds. This will be studied in \cite{50}. There, we
show that the concept of Dirac-Hestenes spinor fields which exists for
$4$-dimensional Lorentzian spin manifolds modelling a relativistic spacetime,
can be generalized for a \textit{\ }the case of general \textit{spin} manifold
of dimension $n=p+q$ (equiped with a metric of signature $(p,q)$, only if the
spinor bundle structure P$_{\mathrm{Spin}_{p,q}^{e}}M$ is trivial.}. However,
these objects exist for a four dimensional vector space $V$ equipped with a
metric of Lorentzian signature and this fact makes them very much important
mathematical objects for physical theories. Indeed, as we shall show in
section 5 it is possible to express Dirac equation in a consistent way using
\emph{DHSF }living on Minkowski spacetime. Such equation is called the
Dirac-Hestenes equation (\emph{DHE}). In section 7 we express the Dirac
equation using \emph{ASF}. In section 4 we define Clifford fields and then
\emph{ASF} and \emph{DHSF}. We observe here that our definitions of \emph{ASF}
and \emph{DHSF} as some equivalence classes of pairs $(\Xi_{u},\psi_{\Xi_{u}%
})$, where $\Xi_{u}$ is a \textit{spinorial coframe}\footnote{Take notice that
in this paper the term spinorial (co)frame field (defined below) is related,
but distinct from the concept of a spin (co)frame, which is a section of a
particular principal bundle called the spin (co)frame bundle (see section 4
and \cite{50}. for more details).} field and $\psi_{\Xi_{u}}$ is an
appropriated Clifford field, i.e., a sum of multivector (or multiform) fields
are not the usual ones that can be found in the literature. These definitions
that, of course, come after the definitions of \emph{AS} and \emph{DHS} are
essentially \emph{different} from the definition of spinors given originally
by Chevalley (\cite{3},\cite{4}). There, spinors are \emph{simply} defined as
elements of a minimal ideal carrying a \emph{modular} representation of the
Clifford algebra $\mathcal{C}\ell(V,\mathbf{g)}$ associated to a structure
$(V,\mathbf{g)}$, where $V$ is a real vector space of dimension $n=p+q$ and
$\mathbf{g}$ is a metric of signature $(p,q)$. And, of course, in that books
there is no definition of \emph{DHS}. Concerning \emph{DHS} we mention that
our definition of these objects \textit{is different also} from the originally
given in (\cite{27}-\cite{29})\footnote{The definitions of \emph{AS},
\emph{DHS,} \emph{ASF} and \emph{DHSF} given below are an improvement over a
preliminary tentative of definitions of these objects given in (\cite{49}).
Unfortunately, that paper contains some equivocated results and errors
(besides many misprints), which we correct here and in \cite{50}. We take the
opportunity to apologize for any incovenients and misunderstandings that
\cite{49} may have caused. Some other papers where related (but not
equivalent) material to the one presented in the present paper and in
\cite{50} can be found are (\cite{5}-\cite{12},\cite{14}-\cite{20}%
,\cite{24}-\cite{26},\cite{31}-\cite{39},\cite{41}-\cite{44},\cite{47}%
,\cite{48}).}. In view of these statements a justification for our definitions
must be given and part of section 5 and section 6 are devoted to such an
enterprise. There it is shown that our definitions are the only \textit{ones}
compatible with the \emph{DHE} and the meaning of the Fierz identities
(\cite{13},\cite{17}). We discuss in section 8 some misunderstandings
resulting from the presentations of the standard Dirac equation when written
with covariant Dirac spinors and also some misunderstandings concerning the
\emph{DHE}.\emph{\ }It is important to emphasize here that the definitions of
\emph{ASF}, \emph{DHSF} \ on Minkowski spacetime and of the spin-Dirac
operator given in section 5 although correct are to be considered only as
\emph{preliminaries}. Indeed, these objects can be defined in a truly
\textit{satisfactory} way on a general Riemann-Cartan spacetime \textit{only}
after the introduction of the concepts of the Clifford and the left (and
right) spin-Clifford bundles. Moreover, a comprehensive formulation of Dirac
equation on these manifolds requires a theory of connections acting on
sections of these bundles. This non trivial subject is studied in a
forthcoming paper (\cite{50}). Section 9 present our conclusions. Finally we
recall that our notations and some necessary results for the intelligibility
of the paper are presented in Appendixes A-E. Although the Appendixes contain
known results, we decided to write them for the benefit of the reader, since
the material cannot be found in a single reference. In particular Appendix A
contains some of the `tricks of the trade' necessary to perform quickly
calculations with Clifford algebras. If the reader needs more details
concerning the theory of Clifford algebras and their applications than the
ones provide by the Appendixes, the references (\cite{5},\cite{21}%
,\cite{22},\cite{26},\cite{30},\cite{39},\cite{45},\cite{46}) will certainly
help. A final remark \ is necessary before we start our enterprise: the theory
of the Dirac-Hestenes spinor fields of this (and the sequel paper \cite{50})
does not contradict the standard theory of covariant Dirac spinor fields that
is used by physicists and indeed it will be shown that the standard theory is
no more than a matrix representation of theory described below.

Some few acronyms are used in the present paper (to avoid long sentences) and
they are summarized below for the reader's convenience.

\textit{AS- }Algebraic Spinor

\textit{ASF}- Algebraic Spinor Field

\textit{CDS- }Covariant Dirac Spinor

\textit{DHE}- Dirac-Hestenes Equation

\textit{DHSF- }Dirac-Hestenes Spinor Field

\section{Algebraic Spinors}

This section introduces the algebraic ideas that motivated the theory of
\emph{ASF }(which will be developed with full rigor in \cite{50}), i.e., we
give a precise definition of \emph{AS}. The algebraic side of the theory of
\emph{DHSF,} namely the concept of\emph{\ } \emph{DHS} is given in section 3.
The justification for that definitions will become clear in sections 5 and 6.

\subsection{\hspace{-0.1cm}Geometrical Equivalence of Representation Modules
of Simple Clifford Algebras $\mathcal{C}\ell(V,\mathbf{g})$}

We start with the introduction of some notations and clarification of some subtleties.

(\textbf{i}) In what follows $V$ is a $n$-dimensional vector space over the
real field $\mathbb{R}$. The dual space of $V$ is denoted $V^{*}$. Let
\begin{equation}
\mathbf{g}:V\times V\rightarrow\mathbb{R} \label{1}%
\end{equation}
be a metric of signature $(p,q).$

(\textbf{ii}) Let $\mathrm{SO}(V,\mathbf{g})$ be the group of endomorphisms of
$V$ that preserves $\mathbf{g}$ and the space orientation. This group is
isomorphic to \textrm{SO}$_{p,q}$ (see Appendix C), but there is no natural
isomorphism. We write $\mathrm{SO}(V,\mathbf{g})\simeq$\textrm{SO}$_{p,q}$.
Also, the connected component to the identity is denoted by $\mathrm{SO}%
^{e}(V,\mathbf{g})$ and $\mathrm{SO}^{e}(V,\mathbf{g})\simeq$\textrm{SO}%
$_{p,q}^{e}$. In the case $p=1$, $q=3,$ $\mathrm{SO}^{e}(V,\mathbf{g})$
preserves besides \emph{orientation} also the \emph{time} orientation. In this
paper we are mainly interested in $\mathrm{SO}^{e}(V,\mathbf{g}).$

(\textbf{iii}) We denote by $\mathcal{C}\ell(V,\mathbf{g})$ the Clifford
algebra\footnote{We reserve the notation $\mathbb{R}_{p,q}$ for the Clifford
algebra of the vector space $\mathbb{R}^{n}$ equipped with a metric of
signature $(p,q)$, $p+q=n$. $C\ell(V,\mathbf{g})$ and $\mathbb{R}_{p,q}$ are
isomorphic, but there is no canonical isomorphism. Indeed, an isomorphism can
be exhibit only after we fix an orthonormal basis of $V$.} of $V$ associated
to $(V,\mathbf{g})$ and by $\mathrm{Spin}^{e}(V,\mathbf{g})$ $(\simeq
\mathrm{Spin}_{p,q}^{e})$ the connected component of the spin group
$\mathrm{Spin}(V,\mathbf{g})$ $\simeq\mathrm{Spin}_{p,q}$ (see Appendix C for
the definitions). Let $\mathbf{L}$ denote $2:1$ homomorphism $\mathbf{L}%
:\mathrm{Spin}^{e}(V,\mathbf{g})\rightarrow\mathrm{SO}^{e}(V,\mathbf{g})$,
$u\mapsto\mathbf{L}(u)\equiv\mathbf{L}_{u}$. $\mathrm{Spin}^{e}(V,\mathbf{g})$
acts on $V$ identified as the space of $1$-vectors of $\mathcal{C}%
\ell(V,\mathbf{g})\simeq\mathbb{R}_{p,q}$ through its adjoint representation
in the Clifford algebra $\mathcal{C}\ell(V,\mathbf{g})$ which is related with
the vector representation of $\mathrm{SO}^{e}(V,\mathbf{g})$ as
follows\footnote{$\mathrm{Aut}(C\ell(V,\mathbf{g}))$ denotes the (inner)
automorphisms of $C\ell(V,\mathbf{g})$.}:
\begin{align}
\mathrm{Spin}^{e}(V,g)\text{ }\ni\text{ }u  &  \mapsto\mathrm{Ad}_{u}%
\in\mathrm{Aut}(\mathcal{C}\ell(V,\mathbf{g}))\nonumber\\
\left.  \mathrm{Ad}_{u}\right\vert _{V}  &  :V\rightarrow V,\text{ }%
\mathbf{v}\mapsto u\mathbf{v}u^{-1}\text{ }=\mathbf{L}_{u}\bullet\mathbf{v}.
\label{2bis}%
\end{align}

In Eq.(\ref{2bis}) $\mathbf{L}_{u}\bullet\mathbf{v}$ denotes the standard
action $\mathbf{L}_{u}$ on $\mathbf{v}$ (see Eq.(\ref{2''})) and \ where
identified (without much ado) $\mathbf{L}_{u}\in\mathrm{SO}^{e}(V,\mathbf{g})$
with $\mathbf{L}_{u}\in\mathbf{V\otimes V}^{\ast}$, $\mathbf{g}\left(
\mathbf{L}_{u}\bullet\mathbf{v,L}_{u}\bullet\mathbf{v}\right)  \mathbf{=g}%
\left(  \mathbf{v,v}\right)  $

(\textbf{iv}) We denote by $\mathcal{C}\ell(V,\mathbf{g})$ the Clifford
algebra\footnote{We reserve the notation $\mathbb{R}_{p,q}$ for the Clifford
algebra of the vector space $\mathbb{R}^{n}$ equipped with a metric of
signature $(p,q)$, $p+q=n$. $C\ell(V,\mathbf{g})$ and $\mathbb{R}_{p,q}$ are
isomorphic, but there is no canonical isomorphism. Indeed, an isomorphism can
be exhibit only after we fix an orthonormal basis of $V$.} of $V$ associated
to $(V,\mathbf{g})$ and by $\mathrm{Spin}^{e}(V,\mathbf{g})$ $(\simeq
\mathrm{Spin}_{p,q}^{e})$ the connected component of the spin group
$\mathrm{Spin}(V,\mathbf{g})$ $\simeq\mathrm{Spin}_{p,q}$ (see Appendix C for
the definitions).

(\textbf{v}) Let $\mathcal{B}$ be the set of all oriented and time oriented
orthonormal basis\footnote{We will call the elements of $\mathcal{B}$ (in what
follows) simply by orthonormal basis.} of $V$. \ Choose among the elements of
$\mathcal{B}$ a basis $\mathit{b}_{0}=\{\mathbf{E}_{1}\mathbf{,....,E}%
_{p,}\mathbf{E}_{p+1,}\mathbf{....,E}_{p+q}\}$, hereafter called the fiducial
frame of $V$. With this choice, we define a $1-1$ mapping
\begin{equation}
\Sigma:\mathrm{SO}^{e}(V,\mathbf{g})\rightarrow\mathcal{B}, \label{2a}%
\end{equation}
given by
\begin{equation}
\mathbf{L}_{u}\mathbf{\mapsto}\Sigma(\mathbf{L}_{u}\mathbf{)}\mathbf{\equiv
}\mathbf{\Sigma}_{\mathbf{L}_{u}}=\mathbf{L}\mathit{b}_{0} \label{2'}%
\end{equation}
where $\mathbf{\Sigma}_{\mathbf{Lu}}=\mathbf{L}_{u}\mathit{b}_{0}$ is a short
for $\{\mathbf{e}_{1}\mathbf{,....,e}_{p,}\mathbf{e}_{p+1,}\mathbf{....,e}%
_{p+q}\}\in\mathcal{B}$, such that denoting the action of $\mathbf{L}_{u}$ on
$\mathbf{E}_{i}\in\mathit{b}_{0}$ by $\mathbf{L}_{u}\mathbf{\bullet E}_{i}$ we
have
\begin{equation}
\mathbf{e}_{i}=\mathbf{L}_{u}\mathbf{\bullet E}_{i}\equiv L^{j}\hspace
{0.02cm}_{i}\mathbf{E}_{j}\text{, }i,j=1,2,...,n. \label{2''}%
\end{equation}
In this way, we can identify a given vector basis $\mathit{b}$ of $V$ with the
isometry $\mathbf{L}_{u}$ that takes the fiducial basis $\mathit{b}_{0}$ to
$\mathit{b}$. \ The fiducial basis $\mathit{b}_{0}$ will be also denoted by
$\mathbf{\Sigma}_{\mathbf{L}_{0}}$, where $\mathbf{L}_{0}=e$, \ is the
identity element of $\mathrm{SO}^{e}(V,\mathbf{g})$.

Since the group $\mathrm{SO}^{e}(V,\mathbf{g})$ is \textit{not} simple
connected their elements cannot distinguish between frames whose spatial axes
are \textit{rotated} in relation to the \ fiducial vector frame $\Sigma
_{\mathbf{L}_{0}}$ by multiples of $2\pi$ or by multiples of $4\pi$. For what
follows it is crucial to make such a distinction. This is done by introduction
of the concept of \textit{spinorial frames}.

\begin{definition}
Let $\mathit{b}_{0}\in\mathcal{B}$ \ be a fiducial frame and choose an
arbitrary $u_{0}\in\mathrm{Spin}^{e}(V,\mathbf{g})$. Fix once and for all the
pair $(u_{0},\mathit{b}_{0})$ with $u_{0}=1$ and call it the fiducial
spinorial frame.
\end{definition}

\begin{definition}
\label{spinframe}The space $\mathrm{Spin}^{e}(V,\mathbf{g})\times\mathcal{B}$
$=\{(u,\mathit{b}),u\mathit{b}u^{-1}=u_{0}\mathit{b}_{0}u_{0}^{-1}\}$ will be
called the space of spinorial frames and denoted by $\mathbf{\Theta}$.
\end{definition}

\begin{remark}
It is crucial for what follows to observe here that the definition
\ref{spinframe} implies that a given $b\in\mathcal{B}$ determines two and only
two spinorial frames, namely $(u,\mathit{b})$ and $(-u,\mathit{b})$, since
$\pm u\mathit{b}(\pm u^{-1})=u_{0}\mathit{b}_{0}u_{0}^{-1}$.
\end{remark}

(\textbf{vi) }We now parallel the construction in (\textbf{v}) but replacing
\ $\mathrm{SO}^{e}(V,\mathbf{g})$ by its universal covering group
$\mathrm{Spin}^{e}(V,\mathbf{g})$ and \ $\mathcal{B}$ by $\mathbf{\Theta}$.
Thus, we define the $1-1$ mapping

\bigskip%
\begin{equation}%
\begin{array}
[c]{cccc}%
\Xi: & \mathrm{Spin}^{e}(V,\mathbf{g}) & \rightarrow & \mathbf{\Theta,}\\
& u & \mapsto & \mathbf{\Xi}(u)\mathbf{\mathbf{\equiv}\Xi}_{u}\mathbf{=(}%
u,\mathit{b}\text{),}%
\end{array}
\label{2b}%
\end{equation}
where $u\mathit{b}u^{-1}=\mathit{b}_{0}$ . \ 

The fiducial spinorial frame will be denoted in what follows by $\mathbf{\Xi
}_{0}$. It is obvious from Eq.(\ref{2b}) that \ $\mathbf{\Xi}%
(-u)\mathbf{=\mathbf{\Xi}}_{\mathbf{\mathbf{(-}}u)}\mathbf{=(-}u,\mathit{b}%
$)$\mathbf{\mathbf{\neq\Xi}}_{u}$.

\begin{definition}
The natural right action of $\ a\in$ $\mathrm{Spin}^{e}(V,\mathbf{g})$ denoted
by $\bullet$ on $\mathbf{\Theta}$ is given by
\begin{equation}
a\bullet\mathbf{\Xi}_{u}=a\bullet\left(  u,\mathit{b}\right)  =(ua,Ad_{a^{-1}%
}\mathit{b})=(ua,a^{-1}\mathit{b}a) \label{2bb}%
\end{equation}

\end{definition}

Observe that if $\mathbf{\Xi}_{u%
%TCIMACRO{\U{b4}}%
%BeginExpansion
\acute{}%
%EndExpansion
}\mathbf{=(}u%
%TCIMACRO{\U{b4}}%
%BeginExpansion
\acute{}%
%EndExpansion
,\mathit{b%
%TCIMACRO{\U{b4}}%
%BeginExpansion
\acute{}%
%EndExpansion
})=u%
%TCIMACRO{\U{b4}}%
%BeginExpansion
\acute{}%
%EndExpansion
\bullet\Xi_{0}$ and $\mathbf{\Xi}_{u}\mathbf{=(}u,\mathit{b})=u\bullet\Xi_{0}$
then,%
\[
\mathbf{\Xi}_{u%
%TCIMACRO{\U{b4}}%
%BeginExpansion
\acute{}%
%EndExpansion
}=(u^{-1}u%
%TCIMACRO{\U{b4}}%
%BeginExpansion
\acute{}%
%EndExpansion
)\bullet\mathbf{\Xi}_{u}=(u%
%TCIMACRO{\U{b4}}%
%BeginExpansion
\acute{}%
%EndExpansion
,u%
%TCIMACRO{\U{b4}}%
%BeginExpansion
\acute{}%
%EndExpansion
^{-1}ubu^{-1}u%
%TCIMACRO{\U{b4}}%
%BeginExpansion
\acute{}%
%EndExpansion
)
\]

Note that there is a natural $2-1$ mapping
\begin{equation}
\mathbf{s}:\mathbf{\Theta\rightarrow}\mathcal{B},\qquad\mathbf{\Xi}_{\pm
u}\mapsto\mathit{b}=(\pm u^{-1})\mathit{b}_{0}(\pm u), \label{2b1}%
\end{equation}
such that
\begin{equation}
\mathbf{s(}(u^{-1}u%
%TCIMACRO{\U{b4}}%
%BeginExpansion
\acute{}%
%EndExpansion
)\bullet\mathbf{\Xi}_{u}))=\mathrm{Ad}_{(u^{-1}u%
%TCIMACRO{\U{b4}}%
%BeginExpansion
\acute{}%
%EndExpansion
)^{-1}}(\mathbf{s}(\mathbf{\Xi}_{u})). \label{2b2}%
\end{equation}
Indeed, $\mathbf{s(}(u^{-1}u%
%TCIMACRO{\U{b4}}%
%BeginExpansion
\acute{}%
%EndExpansion
)\bullet\mathbf{\Xi}_{u}))=\mathbf{s}((u^{-1}u%
%TCIMACRO{\U{b4}}%
%BeginExpansion
\acute{}%
%EndExpansion
)\bullet\left(  u,\mathit{b}\right)  )=u^{\prime-1}u\mathit{b}(u^{\prime
-1}u)^{-1}=\mathit{b%
%TCIMACRO{\U{b4}}%
%BeginExpansion
\acute{}%
%EndExpansion
}=\mathrm{Ad}_{(u^{-1}u%
%TCIMACRO{\U{b4}}%
%BeginExpansion
\acute{}%
%EndExpansion
)^{-1}}\mathit{b}=\mathrm{Ad}_{(u^{-1}u%
%TCIMACRO{\U{b4}}%
%BeginExpansion
\acute{}%
%EndExpansion
)^{-1}}(\mathbf{s}(\mathbf{\Xi}_{u}))$. This means that the natural right
actions of $\mathrm{Spin}^{e}(V,\mathbf{g})$, respectively on $\mathbf{\Theta
}$ and $\mathcal{B}$, commute. In particular, this implies that the spinorial
frames $\mathbf{\Xi}_{u},\mathbf{\Xi}_{-u}\in\mathbf{\Theta}$, which are, of
course distinct, determine the same vector frame $\Sigma_{\mathbf{L}_{u}%
}=\mathbf{s}(\mathbf{\Xi}_{u})=\mathbf{s}(\mathbf{\Xi}_{-u})=\Sigma
_{\mathbf{L}_{-u}}$. We have,%
\begin{equation}
\Sigma_{\mathbf{L}_{u}}=\Sigma_{\mathbf{L}_{-u}}=\mathbf{L}_{u^{-1}u_{0}%
}\Sigma_{\mathbf{L}_{u_{0}}},\text{ }\mathbf{L}_{u^{-1}u_{0}}\in
\mathrm{SO}_{p,q}^{e}, \label{6}%
\end{equation}

Also, from Eq.(\ref{2b2}), we can write explicitly
\begin{equation}
u_{0}\Sigma_{\mathbf{L}_{u_{0}}}u_{0}^{-1}=u\Sigma_{\mathbf{L}_{u}}%
u^{-1},\text{ }u_{0}\Sigma_{\mathbf{L}_{u_{0}}}u_{0}^{-1}=(-u)\Sigma
_{\mathbf{L}_{-u}}(-u)^{-1},u\in\mathrm{Spin}^{e}(V,\mathbf{g}), \label{5}%
\end{equation}
where the meaning of Eq.(\ref{5}) of course, is that \ \textit{if}
$\Sigma_{\mathbf{L}_{u}}=\Sigma_{\mathbf{L}_{-u}}=\mathit{b}=\{\mathbf{e}%
_{1}\mathbf{,....,e}_{p,}\mathbf{e}_{p+1,}\mathbf{....,e}_{q}\}\in\mathcal{B}$
and $\Sigma_{\mathbf{L}_{u_{0}}}=\mathit{b}_{0}\in\mathcal{B}$ is the fiducial
frame, then%
\begin{equation}
u_{0}\mathbf{E}_{j}u_{0}^{-1}=(\pm u)\mathbf{e}_{j}(\pm u^{-1}). \label{3}%
\end{equation}

In resume we can say that the space $\mathbf{\Theta}$ of spinorial frames can
be thought as an \textit{extension} of the space space $\mathcal{B}$ of
\textit{vector frames}, where even if two vector frames have the \textit{same}
ordered vectors, they are considered distinct if the spatial axes of one
vector frame is rotated by a odd number of $2\pi$ rotations relative to the
other vector frame and are considered the same if the spatial axes of one
vector frame is rotated by an even number of $2\pi$ rotations relative to the
other frame. Even if this construction seems to be impossible at first sight,
Aharonov and Susskind \cite{1a} warrants that it can be implemented physically.

(\textbf{vii}) Before we proceed an important \textit{digression} on our
notation used below is necessary. We recalled in appendix B how to construct a
minimum left (or right) ideal for a given real Clifford algebra once a vector
basis $\mathit{b}\in\mathcal{B}$ for $V\hookrightarrow\mathcal{C}%
\ell(V,\mathbf{g})$ is given. That construction suggests to \textit{label } a
given primitive idempotent and its corresponding ideal with the subindex $b$.
However, taking into account the above discussion of vector and spinorial
frames and their relationship\ we find useful for what follows (specially in
view of the definition \ref{isoideal} and the definitions of algebraic and
Dirac-Hestenes spinors (see definitions \ref{aspin} and \ref{dirac-hestenes}
below)) to label a given primitive idempotent and its corresponding ideal with
the subindex $\mathbf{\Xi}_{u}$. Recall after all, that a given idempotent is
according to definition \ref{aspin} \textit{representative} of a particular
spinor in a given spinorial frame $\mathbf{\Xi}_{u}$.

(\textbf{viii) }Next we recall Theorem 49 of Appendix B which says that a
minimal left ideal of $\mathcal{C}\ell(V,\mathbf{g})$ is of the type
\begin{equation}
I_{\mathbf{\Xi}_{u}}=\mathcal{C}\ell(V,\mathbf{g})\mathrm{e}_{\mathbf{\Xi}%
_{u}} \label{7}%
\end{equation}
where $e_{\Xi_{u}}$ is a primitive idempotent of $\mathcal{C}\ell
(V,\mathbf{g})$.

It is easy to see that all ideals $I_{\mathbf{\Xi}_{u}}=\mathcal{C}%
\ell(V,\mathbf{g})\mathrm{e}_{\Xi_{u}}$ and $I_{\mathbf{\Xi}_{u^{\prime}}%
}=\mathcal{C}\ell(V,\mathbf{g})\mathrm{e}_{\mathbf{\Xi}_{u^{\prime}}}$ such
that
\begin{equation}
\mathrm{e}_{\mathbf{\Xi}_{u^{\prime}}}=(u^{\prime-1}u)\mathrm{e}_{\mathbf{\Xi
}_{u}}(u^{\prime-1}u)^{-1} \label{8}%
\end{equation}
$u,u^{\prime}\in\mathrm{Spin}^{e}(V,g)$ are isomorphic. We have the

\begin{definition}
\label{isoideal}Any two ideals $I_{\mathbf{\Xi}_{u}}=\mathcal{C}%
\ell(V,\mathbf{g})\mathrm{e}_{\mathbf{\Xi}_{u}}$ and $I_{\mathbf{\Xi
}_{u^{\prime}}}=\mathcal{C}\ell(V,\mathbf{g})\mathrm{e}_{\mathbf{\Xi
}_{u^{\prime}}}$ such that their generator idempotents are related by
Eq.(\ref{8}) are said geometrically equivalent.
\end{definition}

But take care, no \textit{equivalence relation} has been defined until now. We
observe moreover that we can write
\begin{equation}
I_{\mathbf{\Xi}_{u^{\prime}}}=I_{\mathbf{\Xi}_{u}}(u^{\prime-1}u)^{-1},
\label{9}%
\end{equation}
a equation that will play a key role in what follows.

\subsection{\bigskip Algebraic Spinors of Type $I_{\mathbf{\Xi}_{u}}$}

Let $\{I_{\mathbf{\Xi}_{u}}\}$ be the set of all ideals geometrically
equivalent to a given minimal $I_{\mathbf{\Xi}_{u_{o}}}$ as defined by
Eq.(\ref{9}). Let be%

\begin{equation}
\mathfrak{T}\text{ }=\{(\mathbf{\Xi}_{u},\Psi_{\mathbf{\Xi}_{u}})\text{
}|\text{ }u\in\mathrm{Spin}^{e}(V,\mathbf{g})\text{, }\mathbf{\Xi}_{u}%
\in\mathbf{\Theta}\text{, }\Psi_{\mathbf{\Xi}_{u}}\in I_{\mathbf{\Xi}_{u}}\}.
\label{10}%
\end{equation}
Let $\mathbf{\Xi}_{u},\mathbf{\Xi}_{u^{\prime}}\in\mathbf{\Theta}$,
$\Psi_{\mathbf{\Xi}_{u}}\in I_{\mathbf{\Xi}_{u}}$, $\Psi_{\mathbf{\Xi
}_{u^{\prime}}}\in I_{\mathbf{\Xi}_{u^{\prime}}}$. We define an equivalence
relation $\mathcal{R}$\ on $\mathfrak{T}$\ by setting
\begin{equation}
(\mathbf{\Xi}_{u},\Psi_{\Xi_{u}})\sim(\mathbf{\Xi}_{u^{\prime}},\Psi
_{\Xi_{u^{\prime}}}) \label{11}%
\end{equation}
if and only if $u\mathbf{s(}\Xi_{u})u^{-1}=u^{\prime}\mathbf{s}(\Xi
_{u^{\prime}})u^{\prime-1}$and
\begin{equation}
\Psi_{\mathbf{\Xi}_{u^{\prime}}}u^{\prime-1}=\Psi_{\mathbf{\Xi}_{u}}u^{-1}.
\label{11bis}%
\end{equation}

\begin{definition}
\label{aspin}An equivalence class
\begin{equation}
\mathbf{\Psi}_{\mathbf{\Xi}_{u}}=[(\mathbf{\Xi}_{u},\Psi_{\mathbf{\Xi}_{u}%
})]\in\mathfrak{T/}\mathcal{R} \label{12}%
\end{equation}
\emph{is called an algebraic spinor of type }$I_{\mathbf{\Xi}_{u}}$\emph{\ for
}$\mathcal{C}\ell(V,\mathbf{g})$\emph{. }$\psi_{\mathbf{\Xi}_{u}}\in
I_{\mathbf{\Xi}_{u}}$\emph{\ is said to be a representative of the algebraic
spinor }$\mathbf{\Psi}_{\mathbf{\Xi}_{u}}$\emph{\ in the spinorial frame
}$\mathbf{\Xi}_{u}$.
\end{definition}

We observe that the pairs $(\mathbf{\Xi}_{u},\Psi_{\mathbf{\Xi}_{u}})$ and
$(\mathbf{\Xi}_{-u},\Psi_{\mathbf{\Xi}_{-u}})=(\mathbf{\Xi}_{-u}%
,-\Psi_{\mathbf{\Xi}_{u}})$ are equivalent, but the pairs $(\mathbf{\Xi}%
_{u},\Psi_{\mathbf{\Xi}_{u}})$ and $(\mathbf{\Xi}_{-u},\Psi_{\mathbf{\Xi}%
_{-u}})=(\mathbf{\Xi}_{-u},\Psi_{\mathbf{\Xi}_{u}})$ are not. This distinction
is \textit{essential} in order to give a structure of linear space (over the
real field) to the set $\mathfrak{T}$. Indeed, a natural linear structure on
$\mathfrak{T}$ is given by
\begin{align}
a[(\mathbf{\Xi}_{u},\Psi_{\mathbf{\Xi}_{u}})]+b[(\mathbf{\Xi}_{u}%
,\Psi_{\mathbf{\Xi}_{u}}^{\prime})]  &  =[(\mathbf{\Xi}_{u},a\Psi
_{\mathbf{\Xi}_{u}})]+[(\mathbf{\Xi}_{u^{\prime}},b\Psi_{\mathbf{\Xi}_{u}%
}^{\prime})],\nonumber\\
(a+b)[(\mathbf{\Xi}_{u},\Psi_{\mathbf{\Xi}_{u}})]  &  =a[(\mathbf{\Xi}%
_{u},\Psi_{\mathbf{\Xi}_{u}})]+b[(\mathbf{\Xi}_{u},\Psi_{\mathbf{\Xi}_{u}})].
\label{13}%
\end{align}

The definition that \ we just given is not a standard one in the literature
(\cite{3},\cite{4}). However, the fact is that the standard definition (licit
as it is from the mathematical point of view) is \emph{not} adequate for a
comprehensive formulation of the Dirac equation using algebraic spinor fields
or Dirac-Hestenes spinor fields as we show in a preliminary way in section 5
and in a rigorous and definitive way in a sequel paper \cite{50}.

As observed on Appendix D a given Clifford algebra $\mathbb{R}_{p,q}$ may have
minimal ideals that are not geometrically equivalent since they may be
generated by primitive idempotents that are related by elements of the group
$\mathbb{R}_{p,q}^{\star}$ which are not elements of $\mathrm{Spin}%
^{e}(V,\mathbf{g})$ (see Appendix C where different, non geometrically
equivalent primitive ideals for $\mathbb{R}_{1,3}$ are shown). These ideals
may be said to be of different \emph{types}. However, from the point of view
of the representation theory of the real Clifford algebras (Appendix B) all
these primitive ideals carry equivalent (i.e., isomorphic) \emph{modular}
representations of the Clifford algebra and no preference may be given to any
one\footnote{The fact that there are ideals that are algebraically, but not
geometrically \ equivalent seems to contain the seed for new Physics, see
(\cite{41mosna1},\cite{41MOSNA2}).}. In what follows, when no confusion arises
and the ideal $I_{\mathbf{\Xi}_{u}}$ is clear from the context, we use the
wording algebraic spinor for any one of the possible types of ideals.

\begin{remark}
We observe here that the idea of defintion of algebraic spinor fields as
equivalent classes has it seed in a paper by M. Riez \cite{48a}. However, Riez
used in his defintion simply orthonormal frames instead of the spinorial
frames of our approach. As such, Riez defintion generates contradicitons, as
it is obvious from our discussion above.
\end{remark}

\subsection{\bigskip Algebraic Dirac Spinors}

These are the algebraic spinors associated with the Clifford algebra
$\mathcal{C}\ell(\mathcal{M})\simeq\mathbb{R}_{1,3}$ (the spacetime algebra)
of Minkowski spacetime $\mathcal{M=(}V,\mathbf{\eta}\mathcal{)}$, where
\emph{V} is a four dimensional vector space over $\mathbb{R}$ and
$\mathbf{\eta}$ is a metric of signature $(1,3)$.

Some special features of this important case are:

(a) The group \textrm{Spin}$^{e}(\mathcal{M})$ is the universal covering of
$\mathcal{L}_{+}^{\uparrow}$, the special and orthochronous Lorentz group that
is isomorphic to the group \textrm{SO}$^{e}(\mathcal{M})$ which preserves
\textit{spacetime} orientation and also the \textit{time} orientation
\cite{40} (see also Appendix B).

(b) \textrm{Spin}$^{e}(\mathcal{M})\subset\mathcal{C}\ell^{0}(\mathcal{M})$,
where $\mathcal{C}\ell^{0}(\mathcal{M})\simeq\mathbb{R}_{1,3}$ is the even
subalgebra of $\mathcal{C}\ell(\mathcal{M})$ and is called the Pauli algebra
(see Appendix C).

The most important property is a coincidence given by Eq.(\ref{14}) below. It
permit us to define a \textit{new} kind of spinors.

\bigskip

\section{Dirac-Hestenes Spinors (\emph{DHS})}

Let $\mathbf{\Xi}_{u}\in\mathbf{\Theta}$ be a spinorial frame for$\mathcal{M}$
such that $\mathbf{s(\mathbf{\Xi}}_{u}\mathbf{)}=\{e_{0},e_{1},e_{2}%
,e_{3}\}\in\mathcal{B}$. Then, it follows from Eq.(D18) of Appendix D that
\begin{equation}
I_{\mathbf{\Xi}_{u}}=\mathcal{C}\ell(\mathcal{M})\mathrm{e}_{\mathbf{\Xi}_{u}%
}=\mathcal{C}\ell^{0}(\mathcal{M})\mathrm{e}_{\mathbf{\Xi}_{u}}, \label{14}%
\end{equation}

if
\begin{equation}
\mathrm{e}_{\mathbf{\Xi}_{u}}=\frac{1}{2}(1+e_{0}). \label{15}%
\end{equation}

Then, each $\Psi_{\mathbf{\Xi}_{u}}\in I_{\mathbf{\Xi}_{u}}$ can be written
as
\begin{equation}
\Psi_{\mathbf{\Xi}_{u}}=\psi_{\mathbf{\Xi}_{u}}\mathrm{e}_{\mathbf{\Xi}_{u}%
},\text{ }\psi_{\mathbf{\Xi}_{u}}\in\mathcal{C}\ell^{0}(\mathcal{M)}
\label{16}%
\end{equation}
%

%TCIMACRO{\TEXTsymbol{>}}%
%BeginExpansion
$>$%
%EndExpansion
From Eq.(\ref{11bis}) we get
\begin{equation}
\psi_{\mathbf{\Xi}_{u^{\prime}}}u^{\prime-1}u\mathrm{e}_{\mathbf{\Xi}_{u}%
}=\psi_{\mathbf{\Xi}_{u}}\mathrm{e}_{\mathbf{\Xi}_{u}},\text{ }\psi
_{\mathbf{\Xi}_{u}},\psi_{\mathbf{\Xi}_{u^{\prime}}}\in\mathcal{C}\ell
^{0}(\mathcal{M}). \label{17}%
\end{equation}

A possible solution for Eq.(\ref{17}) is
\begin{equation}
\psi_{\mathbf{\Xi}_{u^{\prime}}}u^{\prime-1}=\psi_{\mathbf{\Xi}_{u}}u^{-1}.
\label{18}%
\end{equation}

Let $\mathbf{\Theta}\mathbf{\times}\mathcal{C}\ell(\mathcal{M})$ and consider
an equivalence relation $\mathcal{E}$ such that%

\begin{equation}
(\mathbf{\Xi}_{u},\phi_{\mathbf{\Xi}_{u}})\sim(\mathbf{\Xi}_{u^{\prime}}%
,\phi_{\mathbf{\Xi}_{u^{\prime}}})\text{ }(\mathrm{mod}\text{ }\mathcal{E)}
\label{19}%
\end{equation}
if and only if $\psi_{\mathbf{\Xi}_{u^{\prime}}}$\ and $\psi_{\mathbf{\Xi}%
_{u}}$\ are related by
\begin{equation}
\phi_{\mathbf{\Xi}_{u^{\prime}}}u^{\prime-1}=\phi_{\mathbf{\Xi}_{u}}u^{-1}.
\label{20}%
\end{equation}
This suggests the following

\begin{definition}
\label{dirac-hestenes}The equivalence classes $[(\mathbf{\Xi}_{u}%
,\phi_{\mathbf{\Xi}_{u}})]\in(\mathbf{\Theta\times}\mathcal{C}\ell
(\mathcal{M}))/\mathcal{E}$\ are the Hestenes spinors. Among the Hestenes
spinors, an important subset is the one consisted of Dirac-Hestenes spinors
where $[(\mathbf{\Xi}_{u},\psi_{\mathbf{\Xi}_{u}})]\in(\mathbf{\Theta\times
}\mathcal{C}\ell^{0}(\mathcal{M}))/\mathcal{E}$. We say that $\phi
_{\mathbf{\Xi}_{u}}$\ $(\psi_{\mathbf{\Xi}_{u}})$\ is a representative of a
Hestenes (Dirac-Hestenes) spinor in the spinorial frame $\mathbf{\Xi}_{u}%
$.\medskip
\end{definition}

How to justify the above definitions of algebraic and Dirac-Hestenes spinors?
The question is answered in the next section.

\section{Clifford Fields, \emph{ASF} and \emph{DHSF}}

The objective of this section is to introduce the concepts of Dirac-Hestenes
spinor fields (\emph{DHSF}) and algebraic spinor fields (\emph{ASF}) living on
Minkowski spacetime. A definitive theory of these objects that can be applied
for arbitrary Riemann-Cartan spacetimes can be given \textit{only} after the
introduction of the Clifford and \ left ( and right) spin-Clifford bundles and
the theory of connections acting on these bundles. This theory will be
presented in \cite{50} and the presentation given below (which can be followed
by readers that have only a rudimentary knowledge of the theory of fiber
bundles) must be considered as a \textit{preliminary} one.

Let $(M,\eta,\tau,\uparrow,\nabla)$ be Minkowski spacetime, where $M$ is
diffeomorphic to $\mathbb{R}^{4}$, $\eta$ is a constant metric field, $\nabla$
is the Levi-Civita connection of $\eta$. $M$ is oriented by $\mathbf{\tau}%
\in\sec\bigwedge\nolimits^{4}M$ and is also time oriented by $\uparrow
$(\cite{53},\cite{54},\cite{56}).

Let\footnote{$\mathrm{P}_{\mathrm{SO}_{1,3}^{e}}M$ is the orthonormal frame
bundle, $\sec\mathrm{P}_{\mathrm{SO}_{1,3}^{e}}M$ means a section of the frame
bundle.} $\{e_{a}\}\in\sec\mathrm{P}_{\mathrm{SO}_{1,3}^{e}}M$ be an
orthonormal (moving) frame\footnote{Orthonormal moving frames are not to be
confused with the concept of \emph{reference frames }The concepts are related,
but distinct. (\cite{53},\cite{54},\cite{56})}, not necessarily a coordinate
frame and let $\gamma^{a}\in\sec T^{\ast}M$ $(a=0,1,2,3)$ be such that the set
$\{\gamma^{a}\}$ is dual to the set $\{e_{a}\}$, i.e., $\gamma^{a}%
(e_{b})=\delta_{b}^{a}.$

The set $\{\gamma^{a}\}$ will be called also a (moving) frame. Let $\gamma
_{a}=\eta_{ab}\gamma^{b}$ , $a,b=0,1,2,3$. The set $\{\gamma_{a}\}$ will be
called the reciprocal frame to the frame $\{\gamma^{a}\}$. Recall
that\footnote{$\check{\eta}$ is the metric of the contangent space and
$\check{\eta}(\gamma^{a},\gamma^{b})=\eta^{ab}=\eta_{ab}=diag(1,-1,-1,-1)$.}
$(T_{x}^{\ast}M,\check{\eta})\simeq\mathcal{M}$. We will denote $(T_{x}^{\ast
}M,\check{\eta})$ by $\mathcal{M}^{\ast}$. Now, due to the \textit{affine}
structure of Minkowski spacetime we can \emph{identify} all the cotangent
spaces as usual. Consider then the Clifford algebra $\mathcal{C}%
\ell(\mathcal{M}^{\ast})$ generated by the coframe $\{\gamma^{a}\},$ where now
we can take $\gamma^{a}:x\mapsto\bigwedge\nolimits^{1}(\mathcal{M}^{\ast
})\subset Cl(\mathcal{M}^{\ast})$ . We have
\begin{equation}
\gamma^{a}(x)\gamma^{b}(x)+\gamma^{b}(x)\gamma^{a}(x)=2\eta^{ab},\text{
}\forall x\in M. \label{21}%
\end{equation}

\begin{definition}
\emph{(preliminary)} A Clifford field is a mapping
\end{definition}

\begin{equation}
C:M\text{ }\ni\text{ }x\mapsto C(x)\in\mathcal{C}\ell(\mathcal{M}^{\ast}).
\label{22}%
\end{equation}
\medskip

In a coframe $\{\gamma^{a}\}$ the expression of a Clifford field is
\begin{equation}
\mathcal{C}=S+A_{a}\gamma^{a}+\frac{1}{2!}B_{ab}\gamma^{a}\gamma^{b}+\frac
{1}{3!}T_{abc}\gamma^{a}\gamma^{b}\gamma^{c}+P\gamma^{5}, \label{23}%
\end{equation}
where $S,A_{a},B_{ab},T_{abc},P$ are scalar functions (the ones with two or
more indices antisymmetric on that indices) and $\gamma^{5}=\gamma^{0}%
\gamma^{1}\gamma^{2}\gamma^{3}$ is the volume element. Saying with other
words, a Clifford field is a sum of non homogeneous differential
forms.\footnote{This result follows once we recall that as a vector space the
Clifford algebra $C\ell(\mathcal{M}^{*})$ is isomorphic to the the Grassmann
algebra $\bigwedge(V^{*})=\sum\limits_{p=0}^{4}\bigwedge^{p}(V^{*})$, where
$\bigwedge^{p}(V^{*})$ is the space of $p$-forms. This is clear from the
definition of Clifford algebra given in the Appendix A. Recall that
$\mathcal{M}^{*}=(V^{*}\simeq T^{*}M,\check{\eta}).$}

Here is the point where a minimum knowledge of the theory of fiber bundles is
required. Minkowski spacetime is parallelizable and admits a \textit{spin
structure}. See, e.g., (\cite{20G},\cite{42}-\cite{44p}, \cite{50}). This
means that Minkowski spacetime has a spin strucutre, i.e., there exists a
principal bundle called the \textit{spin frame bundle} and denoted by
$\mathrm{P}_{\mathrm{Spin}_{1,3}^{e}}M$ that is the double covering of
$\mathrm{P}_{\mathrm{SO}_{1,3}^{e}}M$, i.e., there is a $2:1$ mapping
$\rho:\mathrm{P}_{\mathrm{Spin}_{1,3}^{e}}M\rightarrow\mathrm{P}%
_{\mathrm{SO}_{1,3}^{e}}M$. The elements of $\mathrm{P}_{\mathrm{Spin}%
_{1,3}^{e}}M$ are called the spin frame fields\footnote{When there is no
possibility of confusion we abreviate spin frame field simply as spin frame.},
and if $\digamma_{u}\in\mathrm{P}_{\mathrm{Spin}_{1,3}^{e}}M$ then
$\rho(\digamma_{u})$ $=\{e_{a}\}\in$ $\mathrm{P}_{\mathrm{SO}_{1,3}^{e}}%
M$\ (once we fix a spin frame and associate it to an arbitrary but fixed
element of $\mathrm{u}\in\mathrm{P}_{\mathrm{Spin}_{1,3}^{e}}M$). This means,
that as in section 1, we distinguish frames that differs from a $2\pi$
rotations. Besides $\mathrm{P}_{\mathrm{SO}_{1,3}^{e}}M,$ we introduce also
$\mathrm{P}_{\mathrm{SO}_{1,3}^{e}}^{\prime}M$, the \textit{coframe}
orthonormal bundle, such that for $\{\gamma^{a}\}\in\mathrm{P}_{\mathrm{SO}%
_{1,3}^{e}}^{\prime}M$ there exists $\{e_{a}\}\in$ $\mathrm{P}_{\mathrm{SO}%
_{1,3}^{e}}M$, such that $\gamma^{a}(e_{b})=\delta_{b}^{a}$. Note that
$\{\gamma^{a}\}\in\mathrm{P}_{\mathrm{SO}_{1,3}^{e}}^{\prime}M$, but, as
already observed, take in mind that each $\gamma^{a}:x\mapsto\bigwedge
\nolimits^{1}(\mathcal{M}^{\ast})\subset Cl(\mathcal{M}^{\ast})$. To proceed
choose a fiducial coframe $\{\Gamma^{a}\}\in\mathrm{P}_{\mathrm{SO}_{1,3}^{e}%
}^{\prime}M$, dual to a fiducial frame $\rho(\digamma_{u_{0}})\equiv
\{E_{a}\}\in\sec\mathrm{P}_{\mathrm{SO}_{1,3}^{e}}M$.

Now, let be
\begin{equation}
u:x\mapsto u(x)\in\mathrm{Spin}^{e}(\mathcal{M}^{\ast})\subset\mathcal{C}%
\ell^{0}(\mathcal{M}^{\ast}). \label{24}%
\end{equation}
In complete \textit{analogy} with section 1 let \ $\mathbf{\Theta
}_{\mathcal{M}}^{\prime}=\mathrm{Spin}^{e}(\mathcal{M}^{\ast})\times
\mathrm{P}_{\mathrm{SO}_{1,3}^{e}}^{\prime}M$ be the space of \textit{
spinorial coframe fields. }We define also the $1-1$ mapping
\begin{align}
\Xi &  :\mathrm{Spin}^{e}(\mathcal{M}^{\ast})\rightarrow\mathbf{\Theta
}_{\mathcal{M}}^{\prime}\nonumber\\
u  &  \mapsto\Xi(u)\equiv\Xi_{u}=(u,\{u^{-1}\Gamma_{a}u\}), \label{24bis}%
\end{align}

Note that there is a $2-1$ natural mapping mapping%
\begin{align}
\mathbf{s}^{\prime}  &  :\mathbf{\Theta}_{\mathcal{M}}^{\prime}\ni\Xi
_{u}\mapsto\{\gamma^{a}\}\in\mathrm{P}_{\mathrm{SO}_{1,3}^{e}}^{\prime
}M,\nonumber\\
\gamma^{a}  &  =u^{-1}\Gamma^{a}u \label{24biss}%
\end{align}
Also, denoting the action of $\ \mathrm{a}(x)\in\mathrm{Spin}^{e}%
(\mathcal{M}^{\ast})$ on $\mathbf{\Theta}_{\mathcal{M}}^{\prime}$ by
$\mathrm{a}\bullet\Xi_{u}=(u\mathrm{a},\{\gamma^{a}\})$ we have%
\begin{equation}
\mathbf{\Xi}_{u%
%TCIMACRO{\U{b4}}%
%BeginExpansion
\acute{}%
%EndExpansion
}=(u^{-1}u%
%TCIMACRO{\U{b4}}%
%BeginExpansion
\acute{}%
%EndExpansion
)\bullet\mathbf{\Xi}_{u} \label{24bica}%
\end{equation}%
\begin{equation}
\text{ }\mathbf{s}^{\prime}\mathbf{(}(u^{-1}u%
%TCIMACRO{\U{b4}}%
%BeginExpansion
\acute{}%
%EndExpansion
)\bullet\mathbf{\Xi}_{u}))=\mathrm{Ad}_{(u^{-1}u%
%TCIMACRO{\U{b4}}%
%BeginExpansion
\acute{}%
%EndExpansion
)^{-1}}(\mathbf{s}^{\prime}(\mathbf{\Xi}_{u})). \label{24bisss}%
\end{equation}
As, in the previous section we have associate $1$ $\in$ $\mathrm{Spin}%
^{e}(\mathcal{M}^{\ast})$) to the fiducial \ spinorial coframe field, but of
course we could associated any other element $u_{0};x\mapsto u_{0}(x)$ $\in$
$\mathrm{Spin}^{e}(\mathcal{M}^{\ast})$ to the fiducial spinorial coframe. In
this general case, writing $\Xi_{u_{o}}$ for the fiducial spinorial coframe,
we have $\mathbf{s}^{\prime}(\Xi_{u_{0}})=\{\Gamma^{a}\}$

Note that $\mathbf{s}^{\prime}(\mathbf{\Xi}_{u})=(\mathbf{s}^{\prime
}(\mathbf{\Xi}_{(-u)})$ and that any other coframe field $\mathbf{s}^{\prime
}(\Xi_{u})$ is then related to $\mathbf{s}^{\prime}(\Xi_{u_{o}})$\ by%

\begin{equation}
u_{0}\mathbf{s}^{\prime}(\Xi_{u_{o}})u_{0}^{-1}=\pm u\mathbf{s}^{\prime}%
(\Xi_{u})(\pm u^{-1})=\pm u\mathbf{s}^{\prime}(\Xi_{(-u)})(\pm u^{-1}),
\label{25}%
\end{equation}
where the meaning of this equation is analogous to the one given to
\ Eq.(\ref{5})\medskip, through Eq.(\ref{3})

Taking into account the results of the previous sections and of the Appendices
A, B we are lead to the following definitions.

Let $\{I_{\Xi_{u}}\}$ be the set of all ideals geometrically equivalent to a
given minimal $I_{\Xi_{uo}}$ as defined by Eq.(\ref{9}) where now $u$,
$u^{\prime}$ \emph{are} Clifford fields defined by mappings like the one
defined in Eq.(\ref{24}).

Let be%

\begin{align}
\mathfrak{T}_{\mathcal{M}}\text{ }  &  =\{(x,(\Xi_{u},\Psi_{\Xi_{u}}))\text{
}|\text{ }x\in M\text{, }u(x)\in\mathrm{Spin}^{e}(\mathcal{M}^{\ast})\text{,
}\Xi_{u}\in\mathbf{\Theta}_{\mathcal{M}}^{\prime}\text{,}\nonumber\\
\text{ }\Psi_{\Xi_{u}}  &  :x\mapsto\Psi_{\Xi_{u}}(x)\in I_{\Xi_{u}}\text{,
}\Psi_{\Xi_{u^{\prime}}}:x\mapsto\in\Psi_{\Xi_{u}}(x)\in I_{\Xi_{u^{\prime}}%
}\}. \label{26}%
\end{align}
\hspace{0.5cm}Consider an equivalence relation $\mathcal{R}_{\mathcal{M}}$\ on
$\mathfrak{T}$\ $_{\mathcal{M}}$\ such that
\begin{equation}
(x,(\Xi_{u},\Psi_{\Xi_{u}}))\sim(y,(\Xi_{u^{\prime}},\Psi_{\Xi_{u^{\prime}}}))
\label{27}%
\end{equation}
if and only if $x=y$,
\begin{equation}
u(x)\mathbf{s%
%TCIMACRO{\U{b4}}%
%BeginExpansion
\acute{}%
%EndExpansion
(}\Xi_{u(x)})u^{-1}(x)=u^{\prime}(x)\mathbf{s%
%TCIMACRO{\U{b4}}%
%BeginExpansion
\acute{}%
%EndExpansion
(}\Xi_{u^{\prime}(x)})u^{\prime-1}(x) \label{27bis}%
\end{equation}
and
\begin{equation}
\Psi_{\Xi_{u^{\prime}}}u^{\prime-1}=\Psi_{\Xi_{u}}u^{-1}. \label{28}%
\end{equation}

\begin{definition}
(preliminary) An algebraic spinor field (ASF) of type $I_{\Xi_{u}}$\ for
$\mathcal{M}^{\ast}$is an equivalence class $\mathbf{\Psi}_{\Xi_{u}}%
=[(x,(\Xi_{u},\Psi_{\Xi_{u}}))]\in\mathfrak{T}_{\mathcal{M}}\mathfrak{/}%
\mathcal{R}_{\mathcal{M}}$\medskip. We say that $\Psi_{\Xi_{u}}\in I_{\Xi_{u}%
}$\ is a representative of the ASF $\mathbf{\Psi}_{\Xi_{u}}$\ in the spinorial
coframe field $\Xi_{u}$
\end{definition}

Consider an equivalence relation $\mathcal{E}_{\mathcal{M}}$on the set
$M\times\mathbf{\Xi}_{\mathcal{M}}\times\mathcal{C}\ell(\mathcal{M}^{\ast})$
such that (given $\psi_{\Xi_{u}}:x\mapsto\psi_{\Xi_{u}}(x)\in\mathcal{C}%
\ell(\mathcal{M}^{\ast})$, $\psi_{\Xi_{u^{\prime}}}:x\mapsto\in\psi_{\Xi_{u}%
}(x)\in\mathcal{C}\ell(\mathcal{M}^{\ast})\})$ $((x,(\Xi_{u},\psi_{\Xi_{u}%
})))$ and $((y,(\Xi_{u^{\prime}},\psi_{\Xi_{u^{\prime}}})))$ are equivalent if
and only if $x=y$,%
\begin{equation}
u(x)\mathbf{s}^{\prime}(\Xi_{u(x)})u^{-1}(x)=u^{\prime}(x)\mathbf{s}^{\prime
}(\Xi_{u^{\prime}(x)})u^{\prime-1}(x) \label{29}%
\end{equation}
and%

\begin{equation}
\psi_{\Xi_{u^{\prime}}}u^{\prime-1}=\psi_{\Xi_{u}}u^{-1}. \label{30}%
\end{equation}

\begin{definition}
(preliminary)An equivalence class $\mathbf{\psi}=[(x,(\Xi_{u},\psi_{\Xi_{u}%
}))]\in M\times\mathbf{\Xi}_{\mathcal{M}}\times\mathcal{C}\ell(\mathcal{M}%
^{\ast})/\mathcal{E}_{\mathcal{M}}$ is called a Hestenes spinor field for
$\mathcal{M}^{\ast}$. $\psi_{\Xi_{u}}\in\mathcal{C}\ell(\mathcal{M}^{\ast}%
)$\ is said to be a representative of the Hestenes spinor field $\mathbf{\phi
}_{\Xi_{u}}$\ in the spinorial coframe field $\Xi_{u}$. When $\psi_{\Xi_{u}%
}:x\mapsto\psi_{\Xi_{u}}(x)\in\mathcal{C}\ell^{0}(\mathcal{M}^{\ast})$,
$\psi_{\Xi_{u^{\prime}}}:x\mapsto\in\psi_{\Xi_{u}}(x)\in\mathcal{C}\ell
^{0}(\mathcal{M}^{\ast})$\ we call the equivalence class a Dirac-Hestenes
spinor field (DHSF).
\end{definition}

\section{The Dirac-Hestenes Equation (\emph{DHE})}

In our preliminary presentation of the Dirac equation (on Minkowski spacetime)
that follows we shall restrict our exposition to the case where any spinorial
coframe field appearing in the equations that follows, e.g., $\mathbf{s}%
^{\prime}\mathbf{(}\Xi_{u})=\{\gamma^{a}\}$ is \emph{teleparallel} and
\emph{constant} . By this we mean that $\forall x,y\in M$ and $a=0,1,2,3,$%

\begin{align}
\gamma^{a}(x)  &  \equiv\gamma^{a}(y),\label{32a}\\
\nabla_{e_{a}}\gamma^{b}  &  =0, \label{32b}%
\end{align}
Eq.(\ref{32a}) has meaning due to the \textit{affine }structure of Minkowski
spacetime with permits the usual identification of all tangent spaces (and of
all cotangent spaces) of the manifold and Eq.(\ref{32b}), is the definition of
a teleparallel frame. Of course, the unique solution for Eq.(\ref{32b}) is
\ $\gamma^{\mu}=dx^{\mu}$, where $\{x^{\mu}\}$ are the coordinate functions of
a \textit{global} Lorentz chart of Minkowski spacetime. Such a
\emph{restriction} is a necessary one in our elementary presentation, because
otherwise we would need first to study the theory of the covariant derivative
of spinor fields, a subject that \emph{simply} cannot be appropriately
introduced with the present formalism, thus clearly showing its
\textit{limitation}. Thus, to continue our \textit{elementary} presentation we
need \textit{some} results of the general theory of the covariant derivatives
of spinor fields studied in details in \cite{50}.

Using the results of the previous sections and of the Appendices we can show
(\cite{28},\cite{48WR}) that the \emph{usual} Dirac equation (\cite{1}%
,\cite{15}) (which, as well known is written in terms of covariant Dirac
spinor fields\footnote{Covariant Dirac spinor fields are defined in an obvious
way once we take into account the definition of covariant Dirac spinors given
by Eq.(E6) and Eq.(E7) of the Appendix E. See also (\cite{12},\cite{42}%
-\cite{44}).}) for a representative of a \emph{DHSF} in interaction with an
electromagnetic potential $A:x\mapsto A(x)\in\bigwedge\nolimits^{1}%
(\mathcal{M}^{\ast})\subset\mathcal{C}\ell(\mathcal{M}^{\ast})$ is
\begin{equation}
\mathbf{D}^{s}\psi_{\Xi_{u}}\gamma_{2}\gamma_{1}-m\psi_{\Xi_{u}}\gamma
_{0}+qA\psi_{\Xi_{u}}=0 \label{33}%
\end{equation}

\begin{remark}
It is important for what follows to have in mind that although each
representative $\psi_{\Xi_{u}}:x\mapsto\psi_{\Xi_{u}}(x)\in\mathcal{C}\ell
^{0}(\mathcal{M}^{\ast})$ of a \emph{DHSF} is a sum of nonhomogeneous
differential forms, spinor fields are not a sum of nonhomogeneous differential
forms. Thus, they are mathematical objects of a nature
different\footnote{\textit{Not} taking this difference into account can lead
to misconceptions, as e.g., some appearing in \cite{23}. See our comments in
\cite{52} on that paper.} from that of Clifford fields. The crucial difference
between a Clifford field, e.g., an electromagnetic potential $A$ and a
\emph{DHSF} is that $A$ is frame independent whereas a \emph{DHSF} is frame dependent.
\end{remark}

In the \emph{DHE} the spinor covariant derivative \footnote{If we use more
general frames, that are not Lorentzian coordinate frames, e.g., $\Xi
_{u}=\{\gamma^{a}\}$ then $\mathbf{D}^{s}\psi_{\Xi_{u}}(x)=\gamma^{a}%
\nabla_{e_{a}}^{s}\psi_{\Xi_{u}}(x)=\gamma^{a}(e_{a}+\frac{1}{2}\omega
_{a})\psi_{\Xi_{u}}(x)$, where $\omega_{a}$ is a two form field associated
with the spinorial connection, which is zero \ only for teleparallel frame
fields, if they exist. Details in \cite{50}.} $\mathbf{D}^{s}$ is a first
order differential operator, often called the spin-Dirac operator. Let
$\nabla_{f_{a}}^{s}$ be the spinor covariant derivative. We have the following
representation for $\mathbf{D}^{s}$ in an arbitrary orthonormal frame
$\{t^{a}\}$ dual of the frame $\{f_{a}\}\in\mathrm{P}_{\mathrm{SO}_{1,3}^{e}}%
$.
\begin{equation}
\mathbf{D}^{s}=t^{a}\nabla_{f_{a}}^{s} \label{34a}%
\end{equation}
In a teleparallel spin (co)frame $\mathbf{s}^{\prime}(\Xi_{u})=\{\gamma^{\mu
}\}$ the above equation reduces to
\begin{equation}
\mathbf{D}^{s}=dx^{\mu}\frac{\partial}{\partial x^{\mu}} \label{34}%
\end{equation}

The spin-Dirac operator in an arbitrary orthonormal frame acts on a product
$(\mathcal{C}\psi_{\Xi_{u}})$ where $\mathcal{C}$ is a Clifford field and
$\psi_{\Xi_{u}}$ a representative of a \emph{DHSF }(or a Hestenes field) as a
\emph{modular derivation} (\cite{7},\cite{50}), i.e.,
\[
\mathbf{D}^{s}(C\psi_{\Xi_{u}})=t^{a}\nabla_{f_{a}}^{s}(C\psi_{\Xi_{u}}%
)=t^{a}[(\nabla_{f_{a}}C)\psi_{\Xi_{u}}+C(\nabla_{f_{a}}^{s}\psi_{\Xi_{u}})]
\]

Also in Eq.(\ref{33}) $m$ and $q$ are real parameters (mass and charge)
identifying the elementary fermion described by that equation.\footnote{Note
that we used natural unities in which the value of the velocity of light is
$c=1$ and the value of Planck's constant is $\hbar=1$.}

Now, from Eq.(\ref{30}) we have
\begin{align}
\psi_{\Xi_{u^{\prime}}}  &  =\psi_{\Xi_{u}}s^{-1},\text{ }\Xi_{u%
%TCIMACRO{\U{b4}}%
%BeginExpansion
\acute{}%
%EndExpansion
}=s\bullet\Xi_{u}\label{35}\\
A  &  \mapsto A \label{35B}%
\end{align}
where $s:x\mapsto s(x)\in\mathrm{Spin}^{e}(\mathcal{M}^{\ast})\subset
\mathcal{C}\ell^{0}(\mathcal{M}^{\ast})$ is to be considered a Clifford field.
Consider the case where $s(x)=s(y)=\mathrm{s}$\textrm{, }$\forall x,y\in M$.
Such equation has a precise meaning due to our restriction to teleparallel
frames. We see that the \emph{DHE} is trivially covariant under this kind of
transformation, which can be called a \emph{right} gauge transformation.

Returning to the \emph{DHE} we see also that the equation is
\textit{covariant} under active Lorentz gauge transformations, or \emph{left
}gauge transformations. Indeed, under an active left Lorentz gauge
transformation (\textit{without }changing the spinorial coframe field) we
have,
\begin{align}
\psi_{\Xi_{u}}  &  \mapsto\psi_{\Xi_{u}}^{\prime}=\mathrm{s}\psi_{\Xi_{u}%
},A\mapsto\mathrm{s}A\mathrm{s}^{-1}\nonumber\\
\mathbf{D}^{s}\psi_{\Xi_{u}}  &  \mapsto\mathbf{D%
%TCIMACRO{\U{b4}}%
%BeginExpansion
\acute{}%
%EndExpansion
}^{s}\psi%
%TCIMACRO{\U{b4}}%
%BeginExpansion
\acute{}%
%EndExpansion
_{\Xi_{u}}=\mathrm{s}\mathbf{D}^{s}\psi_{\Xi_{u}}. \label{36}%
\end{align}

The justification for the \ active left Lorentz gauge transformation law
$\mathbf{D}^{s}\psi_{\Xi_{u}}\mapsto\mathbf{D%
%TCIMACRO{\U{b4}}%
%BeginExpansion
\acute{}%
%EndExpansion
}^{s}\psi%
%TCIMACRO{\U{b4}}%
%BeginExpansion
\acute{}%
%EndExpansion
_{\Xi_{u}}=\mathrm{s}\mathbf{D}^{s}\psi_{\Xi_{u}}$ is the
following.\footnote{A study of active \textit{local} left Lorentz gauge
transformations will be presented elsewhere, for it need the concept of
\textit{gauge covariant derivatives.}} The Dirac operator is a 1-form valued
derivative operator $\mathbf{D}^{s}=dx^{\mu}\frac{\partial}{\partial x^{\mu}}%
$. Then, under an active Lorentz gauge transformation $\ \mathrm{s}$ it must
transform like a vector, i.e., $\mathbf{D}^{s}\mapsto\mathbf{D}%
%TCIMACRO{\U{b4}}%
%BeginExpansion
\acute{}%
%EndExpansion
^{s}=\mathrm{s}dx^{\mu}\mathrm{s}^{-1}\frac{\partial}{\partial x^{\mu}}$.

Note that $\psi_{\Xi_{u}}^{\prime}$ is a representative (in the spinorial
coframe field $\Xi_{u}$) of a \emph{new} spinor. Then, it follows, of course,
that the representative of the \emph{new} spinor in the spinorial coframe
field $\Xi_{u^{\prime}}$ is
\begin{equation}
\psi%
%TCIMACRO{\U{b4}}%
%BeginExpansion
\acute{}%
%EndExpansion
_{\Xi_{u%
%TCIMACRO{\U{b4}}%
%BeginExpansion
\acute{}%
%EndExpansion
}}=\mathrm{s}\psi_{\Xi_{u}}\mathrm{s}^{-1}. \label{37}%
\end{equation}
We also recall that the \emph{DHE} is invariant under simultaneous left and
right (\emph{constants}) gauge Lorentz transformations. In this case the
relevant transformations are
\begin{align}
\psi_{\Xi_{u}}  &  \mapsto\psi%
%TCIMACRO{\U{b4}}%
%BeginExpansion
\acute{}%
%EndExpansion
_{\Xi_{u%
%TCIMACRO{\U{b4}}%
%BeginExpansion
\acute{}%
%EndExpansion
}}=\mathrm{s}\psi_{\Xi_{u}}\mathrm{s}^{-1},\nonumber\\
A  &  \mapsto\mathrm{s}A\mathrm{s}^{-1},\text{ \ \ }\mathbf{D}%
%TCIMACRO{\U{b4}}%
%BeginExpansion
\acute{}%
%EndExpansion
^{s}\psi%
%TCIMACRO{\U{b4}}%
%BeginExpansion
\acute{}%
%EndExpansion
_{\Xi_{u%
%TCIMACRO{\U{b4}}%
%BeginExpansion
\acute{}%
%EndExpansion
}}=\mathrm{s}\mathbf{D}^{s}\psi_{\Xi_{u}}\mathrm{s}^{-1}. \label{38}%
\end{align}

\section{Justification of the Transformation Laws of \emph{DHSF} based on the
Fiersz Identities.}

We now give another justification for the definition of Dirac spinors and
\emph{DHSF} presented in previous sections. We start by recalling that a usual
covariant Dirac spinor field determines a set of $p$-form fields, called
bilinear covariants, which describe the physical contents of a particular
solution of the Dirac equation described by that field. The same is true also
for a \emph{DHSF}.

In order to present the bilinear covariants using that fields, we introduce
first the notion of the Hodge dual operator of a Clifford field $\mathcal{C}%
:M$ $\backepsilon$ $x\mapsto\mathcal{C}(x)\in\mathcal{C}\ell(\mathcal{M}%
^{\ast})$. We have

\begin{definition}
The Hodge dual operator is the mapping
\end{definition}

\begin{equation}
\star:\mathcal{C\rightarrow\star C=\tilde{C}}\gamma_{5}, \label{46}%
\end{equation}
\emph{where }$\mathcal{\tilde{C}}$\emph{\ is the reverse of }$\mathcal{C}%
$\emph{\ (}Eq.(A5)\emph{, Appendix A)}.\medskip

Then, in terms of a representative of a \emph{DHSF} in the spinorial frame
field $\Xi_{u}$ the bilinear covariants of Dirac theory reads (with $J=J_{\mu
}\gamma^{\mu}$, $S=\frac{1}{2}S_{\mu\nu}\gamma^{\mu}\gamma^{\nu}$, $K=K_{\mu
}\gamma^{\mu}$)
\begin{equation}%
\begin{array}
[c]{lll}%
\psi_{\Xi_{u}}\tilde{\psi}_{\Xi_{u}}=\sigma+\star\omega &  & \psi_{\Xi_{u}%
}\gamma^{0}\tilde{\psi}_{\Xi_{u}}=J\\
&  & \\
\psi_{\Xi_{u}}\gamma^{1}\gamma^{2}\tilde{\psi}_{\Xi_{u}}=S &  & \psi_{\Xi_{u}%
}\gamma^{0}\gamma^{3}\tilde{\psi}_{\Xi_{u}}=\star S\\
&  & \\
\psi_{\Xi_{u}}\gamma^{3}\tilde{\psi}_{\Xi_{u}}=K &  & \psi_{\Xi_{u}}\gamma
^{0}\gamma^{1}\gamma^{2}\tilde{\psi}_{\Xi_{u}}=\star K\\
&  &
\end{array}
\label{47'}%
\end{equation}

The so called \emph{Fierz identities} are
\begin{equation}
J^{2}=\sigma^{2}+\omega^{2},\hspace{0.5cm}J\cdot K=0,\hspace{0.5cm}%
J^{2}=-K^{2},\hspace{0.5cm}J\wedge K=-(\omega+\star K)S \label{48}%
\end{equation}

\begin{equation}
\left\{
\begin{array}
[c]{lll}%
S\llcorner J=-\omega K &  & S\llcorner K=-\omega J\\
(\star S)\llcorner J=-\sigma K &  & (\star S)\llcorner K=-\sigma J\\
S\cdot S=\sigma^{2}-\omega^{2} &  & (\star S)\cdot S=-2\sigma\omega
\end{array}
\right.  \label{49}%
\end{equation}

\begin{equation}
\left\{
\begin{array}
[c]{l}%
JS=(\omega+\star\sigma)K\\
SJ=-(\omega-\star\sigma)K\\
KS=(\omega+\star\sigma)J\\
SK=-(\omega+\star\sigma)J\\
S^{2}=\omega^{2}-\sigma^{2}+2\sigma(\star\omega)\\
S^{-1}=KSK/J^{4}%
\end{array}
\right.  \label{50}%
\end{equation}

The proof of these identities using the \emph{DHSF} is almost a triviality and
can be done in a few lines. This is not the case if you use covariant Dirac
spinor fields (columns matrix fields). In this case you will need to perform
several pages of matrix algebra calculations.

The importance of the bilinear covariants is due to the fact that we can
recover from them the associate covariant Dirac spinor field (and thus the
\emph{DHSF}) except for a phase. This can be done with an algorithm due to
Crawford \cite{13} and presented in a very pedagogical way in \cite{39}.

Let us consider, e.g., the equation $\psi_{\Xi_{u}}\gamma_{0}\tilde{\psi}%
_{\Xi_{u}}=J$ in (\ref{47'}). Now, $J(x)\in\bigwedge\nolimits^{1}%
(\mathcal{M}^{\ast})\subset\mathcal{C}\ell(\mathcal{M}^{\ast})$ is an
intrinsic object on Minkowski spacetime and according to the accepted first
quantization interpretation theory of the Dirac equation, $qJ$ represents the
electromagnetic current generated by an elementary fermion. The expression of
$J$ in terms of the representative of a \emph{DHSF} in the spinorial coframe
$\Xi_{u^{\prime}}$ is (of course)
\begin{equation}
\psi_{\Xi_{u^{\prime}}}\gamma_{0}^{\prime}\tilde{\psi}_{\Xi_{u^{\prime}}}=J.
\label{51}%
\end{equation}

Now, since
\begin{equation}
\gamma_{0}^{\prime}=(u^{\prime-1}u)\gamma_{0}(u^{\prime-1}u)^{-1}, \label{52}%
\end{equation}
we see that we must have
\begin{equation}
\psi_{\Xi_{u^{\prime}}}=\psi_{\Xi_{u}}(u^{\prime-1}u)^{-1},
\end{equation}
which justifies the definition of \emph{DHSF} given above (see Eq.(\ref{28})).

We observe also that if $\psi_{\Xi_{u}}\tilde{\psi}_{\Xi_{u}}=\sigma
+\star\omega\neq0$, then we can write
\begin{equation}
\psi_{\Xi_{u}}=\rho^{\frac{1}{2}}e^{\frac{1}{2}\beta\gamma^{5}}R, \label{54}%
\end{equation}
where $\forall x\in M,$%
\begin{align}
\rho(x)  &  \in\bigwedge\nolimits^{0}(\mathcal{M}^{\ast})\subset
\mathcal{C}\ell(\mathcal{M}^{\ast})\nonumber\\
\beta(x)  &  \in\bigwedge\nolimits^{0}(\mathcal{M}^{\ast})\subset
\mathcal{C}\ell(\mathcal{M}^{\ast})\label{55}\\
R  &  \in\mathrm{Spin}_{1,3}^{e}(\mathcal{M}^{\ast})\subset\mathcal{C}%
\ell(\mathcal{M}^{\ast})\nonumber
\end{align}

With this result the current $J$ can be written
\begin{equation}
J=\rho v \label{56}%
\end{equation}
with $v=R\gamma^{0}R^{-1}$. Eq.(\ref{56}) discloses the secret geometrical
meaning of \textit{DHSF}. These objects \emph{rotate} and \emph{dilate} vector
fields, this being the reason why they are sometimes called \emph{operator}
spinors (\cite{28}-\cite{30},\cite{39}).

\section{Dirac Equation in Terms of \emph{ASF}}

We recall from Eq.(D2) of Appendix D that
\begin{equation}
\mathrm{e}_{\Xi_{u}}^{\prime}=\frac{1}{2}(1+\gamma_{3}\gamma_{0}) \label{39}%
\end{equation}
is also a primitive idempotent field (here understood as a Hestenes spinor
field) that is algebraically, but not geometrically equivalent to the
idempotent field $\mathrm{e}_{\Xi_{u}}=\frac{1}{2}(1+\gamma_{0})$. Let
$I_{\Xi_{u}}^{\prime}=\mathcal{C}\ell(\mathcal{M}^{\ast})\mathrm{e}_{\Xi_{u}%
}^{\prime}$ be a minimal left ideal generated by $\mathrm{e}_{\Xi_{u}}%
^{\prime}$. Now, multiply the \emph{DHE} (Eq.(\ref{33})) on the left, first by
the primitive idempotent $\mathrm{e}_{\Xi_{u}}$ and then by the primitive
idempotent $\mathrm{e}_{\Xi_{u}}^{\prime}$. We get after some algebra
\begin{equation}
\mathbf{D}^{s}\Phi_{\Xi_{u}}-m\Phi_{\Xi_{u}}(\star1)+qA\Phi_{\Xi_{u}}=0,
\label{40}%
\end{equation}
where $\star1=\gamma_{5}$ is the oriented volume element of Minkowski
spacetime and
\begin{equation}
\Phi_{\Xi_{u}}=\psi_{\Xi_{u}}\mathrm{e}_{\Xi_{u}}\mathrm{e}_{\Xi_{u}}^{\prime
}\in I_{\Xi_{u}}^{\prime}=\mathcal{C}\ell(\mathcal{M}^{\ast})\mathrm{e}%
_{\Xi_{u}}^{\prime} \label{41}%
\end{equation}

Eq.(\ref{40}) is one of the many \emph{faces} of the original equation found
by Dirac in terms of \emph{ASF} and using teleparallel orthonormal frames.

Of course, Eq. (\ref{40}), as it is the case of the \emph{DHE} (Eq.(\ref{33}))
is compatible with the transformation law of \emph{ASF} that follows directly
from the transformation law of \emph{AS} given in section 2. In contrast to
the \emph{DHE, }in Eq.(\ref{40}) there seems to be no explicit reference to
elements of a spinorial coframe field (except for the indices $\Xi_{u}$) since
$\star1$, the volume element is invariant under (Lorentz) gauge
transformations. We emphasize also that the transformation law for \emph{ASF}
is compatible with the presentation of Fierz identities using these objects,
as the interested reader can verify without difficulty.

\section{Misunderstandings Concerning Coordinate Representations of the Dirac
and Dirac-Hestenes Equations.}

We investigate now some \textit{subtleties} of the Dirac and Dirac Hestenes
equations. We start by point out and clarifying some misunderstandings that
often appears in the literature of the subject of the \emph{DHE} when that
equation is presented in terms of a representative of a \emph{DHSF} in a
global coordinate chart $(M,\varphi)$ of the maximal atlas of $M$ with Lorentz
coordinate functions $\langle x^{\mu}\rangle$ associated to it (see, e.g.,
\cite{53}). In that case, $\mathbf{s}^{\prime}(\Xi_{u})=\{\gamma^{\mu}%
=dx^{\mu}\}$. After that we study the (usual) matrix representation of Dirac
equation and show how it hides many features that are only visible in the
\emph{DHE}.

Let $\{e_{\mu}=\frac{\partial}{\partial x^{\mu}}\}$ and $\{e_{\mu}^{\prime
}=\frac{\partial}{\partial x^{\prime\mu}}\}$. The \textit{spinorial} coframe
fields $\Xi_{u}$ and $\Xi_{u^{\prime}}$ (as defined in the previous section)
are associated to the coordinate bases (dual basis) $\mathbf{s}^{\prime}%
(\Xi_{u})=\{\gamma^{\mu}=dx^{\mu}\}$and $\mathbf{s}^{\prime}(\Xi_{u^{\prime}%
})=\{\gamma^{^{\prime}\mu}=dx^{\prime\mu}\}$, corresponding to the global
Lorentz charts $(M,\varphi)$ and $(M,\varphi^{\prime})$. The \emph{DHE} is
written in the charts $\langle x^{\mu}\rangle$ and $\langle x^{\prime\mu
}\rangle$ as
\begin{align}
\gamma^{\mu}(\frac{\partial}{\partial x^{\mu}}\mathbf{\Psi}_{\Xi_{u}}+qA_{\mu
}\mathbf{\Psi}_{\Xi_{u}}\gamma_{1}\gamma_{2})\gamma_{2}\gamma_{1}%
-m\mathbf{\Psi}_{\Xi_{u}}\gamma_{0}  &  =0,\nonumber\\
\gamma^{\prime\mu}(\frac{\partial}{\partial x^{\prime\mu}}\mathbf{\Psi}%
_{\Xi_{u%
%TCIMACRO{\U{b4}}%
%BeginExpansion
\acute{}%
%EndExpansion
}}^{\prime}+qA_{\mu}^{\prime}\mathbf{\Psi}_{\Xi_{u%
%TCIMACRO{\U{b4}}%
%BeginExpansion
\acute{}%
%EndExpansion
}}\gamma_{1}^{\prime}\gamma_{2}^{\prime})\gamma_{2}^{\prime}\gamma_{1}%
^{\prime}-m\mathbf{\Psi}_{\Xi_{u%
%TCIMACRO{\U{b4}}%
%BeginExpansion
\acute{}%
%EndExpansion
}}\gamma_{0}^{\prime}  &  =0, \label{42}%
\end{align}
where $\mathbf{D}^{s}=\gamma^{\mu}\frac{\partial}{\partial x^{\mu}}%
=\gamma^{\prime\mu}\frac{\partial}{\partial x^{\prime\mu}}$ and where
$(\mathbf{\Psi}_{\Xi_{u}},A_{\mu})$ and $(\mathbf{\Psi}_{\Xi_{u^{\prime}}%
},A_{\mu}^{\prime})$ are the coordinate representations of $(\psi_{\Xi_{u}%
},A)$ and $(\psi_{\Xi_{u^{\prime}}},A)$, i.e., for any $x\in M$, we have
\begin{align}
A  &  =A_{\mu}^{^{\prime}}(x^{^{\prime}\mu})dx^{\prime\mu}=A_{\mu}(x^{\mu
})dx^{\mu}\nonumber\\
A_{\mu}^{\prime}(x^{\prime0},x^{\prime1},x^{\prime2},x^{\prime3})  &
=\mathbf{L}_{\mu}^{\nu}A_{\nu}(x^{0},x^{1},x^{2},x^{3}),\nonumber\\
(\mathbf{\Psi}_{\Xi_{u^{\prime}}}U^{\prime-1})|_{(x^{\prime0}(x),x^{\prime
1}(x),x^{\prime2}(x),x^{\prime3}(x))}  &  =(\mathbf{\Psi}_{\Xi_{u}}%
U^{-1})|_{(x^{0}(x),x^{1}(x),x^{2}(x),x^{3}(x))}, \label{43}%
\end{align}
with $U$ and $U^{\prime}$ the coordinate representations of $u$ and
$u^{\prime}$ (see Eq.(\ref{30})) and $L_{\hspace{0.02cm}\mu}^{\nu}$ is an
appropriate Lorentz transformation.

Now, taking into account that the complexification of the algebra
$\mathcal{C}\ell(\mathcal{M}^{\ast})$, i.e., $\mathbb{C\otimes}\mathcal{C}%
\ell(\mathcal{M}^{\ast})$ is isomorphic to the Dirac algebra $\mathbb{R}%
_{4.1}$ (Appendix C), we can think of all the objects appearing in
Eqs.(\ref{42}) as having values also in $\mathbb{C\otimes}\mathcal{C}%
\ell(\mathcal{M}^{\ast})$. Multiply then, both sides of each one of the
Eqs.(\ref{42}) by the following primitive idempotents
fields\footnote{Considered as complexified Hestenes spinor fields (see
Definition 8).} of $\mathbb{C\otimes}\mathcal{C}\ell(\mathcal{M}^{\ast})$ (see
Eq.(D14) of Appendix D)
\begin{align}
f_{\Xi_{u}}  &  =\frac{1}{2}(1+\gamma^{0})\frac{1}{2}(1+i\gamma^{1}\gamma
^{2}),\nonumber\\
f_{\Xi_{u^{\prime}}}  &  =\frac{1}{2}(1+\gamma^{\prime0})\frac{1}{2}%
(1+i\gamma^{\prime1}\gamma^{\prime2}). \label{44}%
\end{align}

Next, look for a matrix representation in $\mathbb{C(}4\mathbb{)}$ of the
resulting equations. We get (using the notation of Appendix D),
\begin{align}
\underline{\gamma}^{\mu}\left(  i\frac{\partial}{\partial x^{\mu}}+qA_{\mu
}(x^{\mu}\mathbf{)}\right)  \mathbf{\Psi(}x^{\mu})-m\mathbf{\Psi(}x^{\mu})  &
=0,\label{45a}\\
\underline{\gamma}^{\mu}\left(  i\frac{\partial}{\partial x^{\prime\mu}%
}+qA_{\mu}^{\prime}(x^{\prime\mu}\mathbf{)}\right)  \mathbf{\Psi}^{\prime
}\mathbf{(}x^{\prime\mu})-m\mathbf{\Psi}^{\prime}\mathbf{(}x^{\prime\mu})  &
=0, \label{45}%
\end{align}
where $\mathbf{\Psi(}x^{\mu}),\mathbf{\Psi}^{\prime}\mathbf{(}x^{\prime\mu})$
are the matrix representations (Eq.(D15), Appendix D) of $\mathbf{\Psi}%
_{\Xi_{u}}$ and $\mathbf{\Psi}_{\Xi_{u^{\prime}}}$. The matrix representations
of the spinors are related by an equation analogous to Eq.(E2) of Appendix E,
\textit{except} that now, these equations refer to fields. The $\{\underline
{\gamma}^{\mu}\}$, $\mu=0,1,2,3$ is the set of Dirac matrices given by
Eq.(D13) of Appendix D. Of course, we arrived at the usual form of the Dirac
equation, \textit{except} for the irrelevant fact that in general the Dirac
spinor is usually represented by a column spinor field, and here we end with a
$4\times4$ matrix field, which however has non null elements only in the first
column.\footnote{The reader can verify without great difficulty that
Eq.(\ref{40}) also has a matrix representation analogous to Eq.(\ref{45}) but
with a set of gamma matrices differing from the set $\{\underline{\gamma}%
^{\mu}\}$ by a similarity transformation.}

Eq.(\ref{45a}), that is the usual presentation of Dirac equation in Physics
textbooks, hides several important facts. First, it hides the basic dependence
of the spinor fields on the \textit{spinorial frame field}, since the
spinorial frames $\Xi_{u}$, $\Xi_{u^{\prime}}$ are such that $\mathbf{s}%
^{\prime}(\Xi_{u})$ $=\{\gamma^{\mu}\}$ and $\mathbf{s}^{\prime}%
(\Xi_{u^{\prime}})=\{\gamma^{\prime\mu}\}$ are mapped on the same set of
matrices, namely $\{\underline{\gamma}^{\mu}\}$. Second, it hides an obvious
geometrical meaning of the theory, as first disclosed by Hestenes
(\cite{28},\cite{29}). Third, taking into account the discussion in a previous
section, we see that the usual presentation of Dirac equation does
\textit{not} leave clear at all if we are talking about \emph{passive} or
\emph{active} Lorentz gauge transformations. Finally, since diffeomorphisms on
the world manifold are in general erroneous associated with coordinate
transformations in many Physics textbooks, Eq.(\ref{45a}) suggests that
spinors must change under diffeomorphisms in a way different from the true
one, for indeed Dirac spinor fields (and also, \textit{DHSF}) are scalars
under diffeomorphisms, an issue that we will discuss in another publication.

\section{Conclusions}

In this paper we investigated how to define algebraic and Dirac-Hestenes
spinor fields on \textit{Minkowski} spacetime. We showed first, that in
general, \textit{algebraic spinors} can be defined for any \textit{\ }real
vector space of any dimension and equipped with a non degenerated metric of
arbitrary signature, but that is not the case for \textit{Dirac-Hestenes
spinors}. These objects exist for a four dimensional real vector space
equipped with a metric of Lorentzian signature. It is this \textit{fact} that
make them very important objects (and gave us the desire to present a rigorous
mathematical theory for them), since as shown in sections 5 and 7 the Dirac
equation can be written in terms of \textit{Dirac-Hestenes spinor}
\textit{fields} or \textit{algebraic spinor fields}. We observe that our
definitions of algebraic and Dirac-Hestenes spinor fields as some equivalence
classes in appropriate sets are \textit{not} the standard ones and the
\textit{core} of the paper was to give genuine motivations for them. We
observe moreover that the definitions of Dirac-Hestenes spinor fields and of
the spin-Dirac operator given in section 5 although correct are to be
considered only as \emph{preliminaries}. The reason is that \textit{any}
rigorous presentation of the theory of the spin-Dirac operator (an in
particular, on a general Riemann-Cartan spacetime) can only be given after the
introduction of the concepts of Clifford and spin-Clifford bundles over these
spacetimes. This is studied in a sequel paper \cite{50}. In \cite{52} we show
some non trivial applications of the concept of Dirac-Hestenes spinor fields
by proving (\textit{mathematical}) Maxwell-Dirac equivalences of the first and
second kinds and showing how these equivalences can eventually put some light
on a possible physical interpretation of the famous Seiberg-Witten equations
for Minkowski spacetime.$\medskip$

\textbf{Noted added in proof }: After we finished the writing of the present
paper and of \ \cite{50}, we learned about the very interesting papers by
Marchuck (\cite{100}-\cite{108}). There, a different point of view concerning
the writing of the Dirac equation using tensor fields\footnote{Paper
\cite{100}, indeed, use a particular case of objects that we called extensors
in a recent series of papers \cite{21,22,22a,22b,22c,22d,22e}.} is developed.
We will discuss Marchuck papers on another place.

\textbf{Acknowledgments}: The author is grateful to Doctors V. V.
Fern\'{a}ndez, A. Jadczyk, P. Lounesto, D. Miralles, A. M. Moya, I. Porteous
and J. Vaz, Jr. and specially to R. A. Mosna for discussions and to F. G.
Rodrigues for help with latex. The author is also grateful grateful to CNPq
(Brazil) for a generous research grant (contract 201560/82-8). Thanks are also
due to the Department of Mathematical Sciences of the University of Liverpool
(where part of this research was done) for providing a Visiting Professor
position and an enjoyable atmosphere for research. The paper is dedicated to
the memory of the late P. Lounesto which collaborated with the author on the
past on the subject of this paper and from whom he learned a lot about the
theory of Cllifford algebras.

\appendix

\section{Some Features about Real and Complex Clifford Algebras}

In this Appendix we fix the notations that we used and introduce the main
ideas concerning the theory of Clifford algebras necessary for the
intelligibility of the paper.

\subsection{Definition of the Clifford Algebra $\mathcal{C}\ell(\mathbf{V,b)}%
$}

In this paper we are interested only in Clifford algebras of a vector
space\footnote{We reserve the notation $V$ for real vector spaces.}
$\mathbf{V}$\textbf{\ }of finite dimension $n$ over a field $\mathbb{F=R}$ or
$\mathbb{C}$. Let $\mathbf{q:V\rightarrow}\mathbb{F}$ be a non degenerate
quadratic form over $\mathbf{V}$ with values in $\mathbb{F}$ and
$\mathbf{b:V\times V\rightarrow}\mathbb{F}$ the associated bilinear form
(which we call a \emph{metric} in the case $\mathbb{F=R}$). We use the
notation
\begin{equation}
x\cdot y=\mathbf{b}(x,y)=\frac{1}{2}(\mathbf{q}(x+y)-\mathbf{q}(x)-\mathbf{q}%
(y) \label{A1}%
\end{equation}

Let $\bigwedge\mathbf{V=}\sum\limits_{i=0}^{n}\bigwedge^{i}\mathbf{V}$ be the
exterior algebra of $\mathbf{V}$ where $\bigwedge^{i}\mathbf{V}$ is
the$\left(  \overset{n}{i}\right)  $ dimensional space of the $i$-vectors.
$\bigwedge^{0}\mathbf{V}$ is identified with $\mathbb{F}$ and $\bigwedge
^{1}\mathbf{V}$ is identified with $\mathbf{V}$. The dimension of
$\bigwedge\mathbf{V}$ is $2^{n}$. A general element $X\in\bigwedge\mathbf{V}$
is called a \emph{multivector} and be written as
\begin{equation}
X=\sum\limits_{i=0}^{n}\langle X\rangle_{i},\text{ }\langle X\rangle_{i}%
\in\bigwedge\nolimits^{i}\mathbf{V,} \label{A2}%
\end{equation}
where
\begin{equation}
\langle\rangle_{i}:\bigwedge\mathbf{V\rightarrow}\bigwedge\nolimits^{i}%
\mathbf{V} \label{A3}%
\end{equation}
is the projector in $\bigwedge\nolimits^{i}\mathbf{V}$, also called the
$i$-part of $X$.

\begin{definition}
The main involution or \emph{grade} involution is an automorphism
\end{definition}

\begin{equation}
\symbol{94}:\bigwedge\mathbf{V\ni X\mapsto\hat{X}\in}\bigwedge\mathbf{V}
\label{A4}%
\end{equation}
$\emph{such}$ $\emph{that}$%

\begin{equation}
\hat{X}=\sum\limits_{k=0}^{n}(-1)^{k}\langle X\rangle_{k}. \label{A4a}%
\end{equation}
$\hat{X}$\emph{\ is called the grade involution of }$X$\emph{\ or simply the
involuted of }$X$\emph{.\medskip}

\begin{definition}
The reversion operator is the anti-automorphism
\end{definition}

\begin{equation}
\symbol{126}:\bigwedge\mathbf{V\ni X\mapsto\tilde{X}\in}\bigwedge\mathbf{V}
\label{A5}%
\end{equation}
\emph{such that}
\begin{equation}
\tilde{X}=\sum\limits_{k=0}^{n}(-1)^{\frac{1}{2}k(k-1)}\langle X\rangle_{k},
\label{A5a}%
\end{equation}
$\tilde{X}$\emph{\ is called the reverse of }$X$.\medskip

The composition of the grade evolution with the reversion operator, denote by
$-$ is called by some authors (e.g., \cite{39},\cite{45},\cite{46}) the
\emph{conjugation} and, $\bar{X}$ is called the \emph{conjugate} of $X$. We
have $\bar{X}=(\tilde{X})=(\hat{X})$.

Since the grade and reversion operators are involutions on the vector space of
multivectors, we have that $\widehat{\hat{X}}=X$ and $\widetilde{\tilde{X}}%
=X$. both involutions commute with the $k$-part operator, i.e., $\widehat
{\langle X\rangle_{k}}=\langle\hat{X}\rangle_{k}$ and $\widetilde{\langle
X\rangle_{k}}=\langle\tilde{X}\rangle_{k}$, for each $k=0,1,...,n$.

\begin{definition}
The exterior product of multivectors $X$ and $Y$ is defined by
\end{definition}

\begin{equation}
\langle X\wedge Y\rangle_{k}=\sum\limits_{j=0}^{k}\langle X\rangle_{j}%
\wedge\langle Y\rangle_{k-j}, \label{A6}%
\end{equation}
\emph{for each }$k=0,1,...,n$\emph{. Note that on the right side there appears
the exterior product}\footnote{We assume that the reader is familiar with the
exterior algebra. We only caution that there are some different definitions of
the exterior product in terms of the tensor product differing by numerical
factors. This may lead to some confusions, if care is not taken. Details can
be found in (\cite{21},\cite{22}).}\emph{\ of }$j$\emph{-vectors and (}%
$k-j$\emph{)-vectors with }$0\leq j\leq n$.\medskip

This exterior product is an \emph{internal }composition law on $\bigwedge
\mathbf{V}$. It is associative and satisfies the distributives laws (on the
left and on the right).

\begin{definition}
The vector space $\bigwedge\mathbf{V}$ endowed with this exterior product
$\wedge$ is an associative algebra called the exterior algebra of multivectors.
\end{definition}

We recall now some of the most important properties of the exterior algebra of multivectors:

For any $\alpha,\beta\in\mathbb{F},$ $X\mathbb{\in}\bigwedge\mathbf{V}$%
\begin{align}
\alpha\wedge\beta &  =\beta\wedge\alpha=\alpha\beta\text{ (product of
}\mathbb{F}\text{ numbers)}\nonumber\\
\alpha\wedge X  &  =X\wedge\alpha=\alpha X\text{ (multiplication by scalars)}
\label{A7}%
\end{align}

For any $X_{j}\in\bigwedge^{j}\mathbf{V}$ and $Y_{k}\in\bigwedge^{k}%
\mathbf{V}$%
\begin{equation}
X_{j}\wedge Y_{k}=(-1)^{jk}Y_{k}\wedge X_{j}. \label{A8}%
\end{equation}

For any $X,Y\in\bigwedge\mathbf{V}$%
\begin{align}
\widehat{X\wedge Y}  &  =\hat{X}\wedge\hat{Y},\nonumber\\
\widetilde{X\wedge Y}  &  =\tilde{X}\wedge\tilde{Y}. \label{A9}%
\end{align}

\subsection{Scalar product of multivectors}

\begin{definition}
A scalar product between the multivectors $X,Y\in\bigwedge\mathbf{V}$ is given by
\end{definition}

\begin{equation}
X\cdot Y=\sum\limits_{i=0}^{n}\langle X\rangle_{i}\cdot\langle Y\rangle_{i},
\label{A10}%
\end{equation}
\emph{where }$\langle X\rangle_{0}\cdot\langle Y\rangle_{0}=\langle
X\rangle_{0}\langle Y\rangle_{0}$\emph{\ is the multiplication in the field
}$\mathbb{F}$\emph{\ and }$\langle X\rangle_{i}\cdot\langle Y\rangle_{i}%
$\emph{\ is given by Eq.(A2), and writing}
\begin{align}
\langle X\rangle_{k}  &  =\frac{1}{k!}X^{i_{1}i_{2}...i_{k}}b_{i_{1}}\wedge
b_{i_{2}}\wedge...\wedge b_{i_{k}},\nonumber\\
\langle Y\rangle_{k}  &  =\frac{1}{k!}Y^{i_{1}i_{2}...i_{k}}b_{i_{1}}\wedge
b_{i_{2}}\wedge...\wedge b_{i_{k}} \label{A11}%
\end{align}
\emph{where }$\left\{  b_{k}\right\}  ,k=1,2,...,n$\emph{\ is an arbitrary
basis of }$\mathbf{V}$\emph{\ we have}
\begin{equation}
\langle X\rangle_{k}\cdot\langle Y\rangle_{k}=\frac{1}{(k!)^{2}}X^{i_{1}%
i_{2}...i_{k}}Y^{j_{1}j_{2}...j_{k}}(b_{i_{1}}\wedge b_{i_{2}}...b_{i_{k}%
})\cdot(b_{j_{1}}\wedge b_{j_{2}}...b_{j_{k}}), \label{A11.a}%
\end{equation}
\emph{with}
\begin{equation}
(b_{i_{1}}\wedge b_{i_{2}}\wedge...\wedge b_{i_{k}})\cdot(b_{j_{1}}\wedge
b_{j_{2}}\wedge...\wedge b_{j_{k}})=\left\vert
\begin{array}
[c]{cccc}%
b_{i_{1}}\cdot b_{j_{1}} & ... & ... & b_{i_{1}}\cdot b_{j_{k}}\\
......... & ... & ... & .........\\
......... & ... & ... & .........\\
b_{i_{k}}\cdot b_{j_{1}} & ... & ... & b_{i_{k}}\cdot b_{j_{k}}%
\end{array}
\right\vert . \label{A12}%
\end{equation}

It is easy to see that for any $X,Y\in\bigwedge\mathbf{V}$%
\begin{align}
\hat{X}\cdot Y  &  =X\cdot\hat{Y},\nonumber\\
\tilde{X}\cdot Y  &  =X\cdot\tilde{Y}. \label{A13}%
\end{align}

\begin{remark}
Observe that the definition of the scalar product given in this paper by
Eq.(A12) differs by a signal from the scalar product of multivectors defined,
e.g., in \cite{27}. Our definition is a \emph{natural} one if we start the
theory with the euclidean Clifford algebra of multivectors of a real vector
space \textbf{V}. The euclidean Clifford algebra is fundamental for the
construction of the theory of extensors and extensor fields (\cite{21}%
,\cite{22},\cite{22a},\cite{22b},\cite{22c},\cite{22d},\cite{22e}).
\end{remark}

\subsection{Interior Algebras}

\begin{definition}
We define two different \emph{contracted products} for arbitrary multivectors
$X,Y\in\bigwedge\mathbf{V}$ by
\end{definition}

\begin{align}
(X\lrcorner Y)\cdot Z  &  =Y(\tilde{X}\wedge Z),\nonumber\\
(X\llcorner Y)  &  =X\cdot(Z\wedge\tilde{Y}), \label{A14}%
\end{align}
where $Z\in\bigwedge\mathbf{V}$.\ $\ $The internal composition rules
$\lrcorner$\emph{\ and }$\llcorner$\emph{\ will be called respectively the
left and the right contracted product. }

These contracted products $\lrcorner$ and $\llcorner$ are internal laws on
$\bigwedge\mathbf{V}$. Both contract products satisfy the distributive laws
(on the left and on the right) but they are \emph{not }associative.

\begin{definition}
The vector space $\bigwedge\mathbf{V}$ endowed with each one of these
contracted products (either $\lrcorner$ or $\llcorner$) is a non-associative
algebra. They are called the interior algebras of multivectors.\medskip
\end{definition}

We present now some of the most important properties of the interior products:

(a) For any $\alpha,\beta\in\mathbb{F},$ and $X\mathbb{\in}\bigwedge
\mathbf{V}$%
\begin{align}
\alpha\lrcorner\beta &  =\alpha\llcorner\beta=\alpha\beta\text{ (product in
}\mathbb{F}\text{),}\nonumber\\
\alpha\lrcorner X  &  =X\llcorner\alpha=\alpha X\text{ (multiplication by
scalars).} \label{A15}%
\end{align}

(b) For any $X_{j}\in\bigwedge^{j}\mathbf{V}$ and $Y_{k}\in\bigwedge
^{k}\mathbf{V}$ with $j\leq k$%
\begin{equation}
X_{j}\lrcorner Y_{k}=(-1)^{j(k-j)}Y_{k}\llcorner X_{j}. \label{A16}%
\end{equation}

(c) For any $X_{j}\in\bigwedge^{j}\mathbf{V}$ and $Y_{k}\in\bigwedge
^{k}\mathbf{V}$%
\begin{align}
X_{j}\lrcorner Y_{k}  &  =0,\text{ if }j>k,\nonumber\\
X_{j}\llcorner Y_{k}  &  =0,\text{ if }j<k. \label{A17}%
\end{align}

(d) For any $X_{k},Y_{k}\in\bigwedge^{k}\mathbf{V}$%
\begin{equation}
X_{j}\lrcorner Y_{k}=X_{j}\llcorner Y_{k}=\tilde{X}_{k}\cdot Y_{k}=X_{k}%
\cdot\tilde{Y}_{k}. \label{A18}%
\end{equation}

(e) For any $v\in\mathbf{V}$ and $X,Y\in\bigwedge\mathbf{V}$%
\begin{equation}
v\lrcorner(X\wedge Y)=(v\lrcorner X)\wedge Y+\hat{X}\wedge(v\lrcorner Y).
\label{A19}%
\end{equation}

\subsection{Clifford Algebra $\mathcal{C}\ell(\mathbf{V,b)}$}

\begin{definition}
The \emph{Clifford product} of multivectors $X$ and $Y$ (denoted by
juxtaposition) is given by the following axiomatic:
\end{definition}

($\emph{i}$) For all $\alpha\in\mathbb{F}$ and $X\in\bigwedge\mathbf{V}:\alpha
X=X\alpha$ equals multiplication of multivector $X$ by scalar $\alpha$.

($\emph{ii}$) For all $v\in\mathbf{V}$ and $X\in\bigwedge\mathbf{V}%
:vX=v\lrcorner X+v\wedge X$ and $Xv=X\llcorner v+X\wedge v$.

($\emph{iii}$) For all $X,Y,Z\in\bigwedge\mathbf{V}:X(YZ)=(XY)Z$.

The Clifford product is an internal law on $\bigwedge\mathbf{V}$. It is
associative (by the axiom ($\emph{iii}$)) and satisfies the distributives laws
(on the left and on the right). The distributive laws follow from the
corresponding distributive laws of the contracted and exterior products.

\begin{definition}
The vector space of multivectors over \textbf{V} endowed with the Clifford
product is an associative algebra with unity called $\mathcal{C}%
\ell(\mathbf{V,b)}$.
\end{definition}

\subsection{Relation Between the Exterior and the Clifford Algebras and the
Tensor Algebra}

Modern algebra books give the

\begin{definition}
The exterior algebra of $\mathbf{V}$ is the quotient algebra $\bigwedge
\mathbf{V=}T(\mathbf{V})/I$, where $T(\mathbf{V})$ is the tensor algebra of
$\mathbf{V}$ and $I\subset T(\mathbf{V})$ is the bilateral ideal generated by
the elements of the form $\mathbf{x\otimes x,}$ $\mathbf{x\in V}$.
\end{definition}

\begin{definition}
The Clifford algebra of $(\mathbf{V},\mathbf{b)}$ is the quotient algebra
$\mathcal{C}\ell(\mathbf{V,b)=}T(\mathbf{V})/I_{b}$, where $I_{b}$ is the
bilateral ideal generated by the elements of the form $\mathbf{x\otimes
x-2b(x,x),}$ $\mathbf{x\in V}$.
\end{definition}

We can show that this definition is equivalent to the one given
above\footnote{When the exterior algebra is defined as $\bigwedge
\mathbf{V=}T(\mathbf{V})/I$ and the Clifford algebra as $\mathcal{C}%
\ell(\mathbf{V,b})=T(\mathbf{V})/I_{\mathbf{b}}$, the (associative) exterior
product of the multivectors in the terms of the tensor product of these
multivectors is fixed once and for all. We have, e.g., that for $x,y\in
\mathbf{V},$ $x\wedge y=\frac{1}{2}(x\otimes y-y\otimes x)$. However, keep in
mind that it possible to define an (associative) exterior product in
$\bigwedge\mathbf{V}$ differing from the above one by numerical factors, and
indeed in (\cite{21},\cite{22},\cite{22a},\cite{22b},\cite{22c},\cite{22d}%
,\cite{22e}) we used another choice. When reading a text on the subject it is
a good idea to have in mind the definition used by the author, for otherwise
confusion may result.}. The space $\mathbf{V}$\textbf{\ }is naturally
\emph{embedded} on $\mathcal{C}\ell(\mathbf{V,b)}$, i.e.,
\begin{align}
\mathbf{V}\overset{i}{\hookrightarrow}T(\mathbf{V})\overset{j}{\rightarrow
}T(\mathbf{V})/I_{\mathbf{b}}  &  =\mathcal{C}\ell(\mathbf{V,b),}\nonumber\\
\text{ and }\mathbf{V}  &  \equiv j\circ i(\mathbf{V})\subset\mathcal{C}%
\ell(\mathbf{V,b).} \label{A20}%
\end{align}
Let $\mathcal{C}\ell^{0}(\mathbf{V,b)}$ and $\mathcal{C}\ell^{1}%
(\mathbf{V,b)}$ be respectively the $j$-images of $\oplus_{i=0}^{\infty}%
T^{2i}(\mathbf{V})$ and $\oplus_{i=0}^{\infty}T^{2i+1}(\mathbf{V})$ in
$\mathcal{C}\ell(\mathbf{V,b)}$. The elements of $\mathcal{C}\ell
^{0}(\mathbf{V,b)}$ form a sub-algebra of $\mathcal{C}\ell(\mathbf{V,b)}$
called the even sub-algebra of $\mathcal{C}\ell(\mathbf{V,b)}$. Also, there is
a canonical vector isomorphism\footnote{The isomorphism is compatible with the
filtrations of the filtered algebra $\bigwedge\mathbb{V}$, i.e., $\left(
\bigwedge^{r}\mathbb{V}\right)  \wedge\left(  \bigwedge^{s}\mathbb{V}\right)
\subseteq\bigwedge^{r+s}\mathbb{V}$.} $\bigwedge\mathbf{V\rightarrow
}\mathcal{C}\ell(\mathbf{V,b)}$, which permits to speak of the embeddings
$\bigwedge^{p}\mathbf{V}\subset\mathcal{C}\ell(\mathbf{V,b)}$, $0\leq p\leq
n$, where $n$ is the dimension of $\mathbf{V}$ (\cite{7}).

\subsection{Some Useful Properties of the Real Clifford Algebras
$\mathcal{C}\ell(V\mathbf{,g)}$}

We now collect some useful formulas which hold for a real Clifford algebra
$\mathcal{C}\ell(V\mathbf{,g)}$ and which has been used in calculations in the
text and Appendices\footnote{As the reader can verify, many of these
properties are also valid for the complex Clifford algebras.}.

For any $v\in V$ and $X\in\bigwedge V$%
\begin{align}
v\lrcorner X  &  =\frac{1}{2}(vX-\bar{X}v)\text{ and }X\lrcorner v=\frac{1}%
{2}(Xv-v\bar{X}),\nonumber\\
v\wedge X  &  =\frac{1}{2}(vX+\bar{X}v)\text{ and }X\wedge v=\frac{1}%
{2}(Xv+v\bar{X}). \label{A21}%
\end{align}

For any $X,Y\in V$%
\begin{equation}
X\cdot Y=\langle\tilde{X}Y\rangle_{0}=\langle X\tilde{Y}\rangle_{0}.
\label{A22}%
\end{equation}

For any $X,Y,Z\in V$%
\begin{align}
(XY)\cdot Z  &  =Y\cdot(\tilde{X}Z)=X\cdot(Z\tilde{Y}),\nonumber\\
X\cdot(YZ)  &  =(\tilde{Y}X)\cdot Z=(X\tilde{Z})\cdot Y. \label{A23}%
\end{align}

For any $X,Y\in V$%
\begin{align}
\overline{XY}  &  =\bar{X}\bar{Y},\nonumber\\
\widetilde{XY}  &  =\tilde{Y}\tilde{X}. \label{A24}%
\end{align}

Let $I\in\bigwedge^{n}V$ then for any $v\in V$ and $X\in\bigwedge V$%
\begin{equation}
I(v\wedge X)=(-1)^{n-1}v\lrcorner(IX). \label{A25}%
\end{equation}

Eq.(A22) is sometimes called the \emph{duality} identity and plays an
important role in the applications involving the Hodge dual operator (see
Eq.(\ref{46})).

For any $X,Y,Z\in V$%
\begin{align}
X\lrcorner(Y\wedge Z)  &  =(X\wedge Y)\lrcorner Z,\nonumber\\
(X\llcorner Y)\llcorner Z  &  =X\llcorner(Y\wedge Z). \label{A26}%
\end{align}

For any $X,Y\in V$%
\begin{equation}
X\cdot Y=\langle\tilde{X}Y\rangle_{0} \label{A27}%
\end{equation}

For\footnote{We observe also that when $K=\mathbb{R}$ and the quadratic form
is euclidean then $X\cdot Y$ is positive definite.} $X_{r}\in\bigwedge^{r}V,$
$Y_{s}\in\bigwedge^{s}V$ we have
\begin{equation}
X_{r}Y_{s}=\langle X_{r}Y_{s}\rangle_{\left\vert r-s\right\vert }+\langle
X_{r}Y_{s}\rangle_{\left\vert r-s\right\vert +2}+...+\langle X_{r}Y_{s}%
\rangle_{r+s}. \label{A28}%
\end{equation}

\section{Representation Theory of the Real Clifford Algebras $\mathbb{R}%
_{p,q}$}

The real Clifford algebras $\mathbb{R}_{p,q}$ are associative algebras and
they are simple or semi-simple algebras. For the intelligibility of the
present paper, it is then necessary to have in mind some results concerning
the presentation theory of associative algebras, which we collect in what
follows, without presenting proofs.

\subsection{Some Results from the Representation Theory of Associative
Algebras.}

Let $\mathbf{V}$ be a set and $\mathbb{K}$ a division ring. Give to the set
$\mathbf{V}$ a structure of \emph{finite} dimensional linear space over
$\mathbb{K}$. Suppose that $\dim_{\mathbb{K}}\mathbf{V}=n$, where
$n\in\mathbb{Z}$. We are interested in what follows in the cases where
$\mathbb{K=R},$ $\mathbb{C}$ or $\mathbb{H}$. When $\mathbb{K=R},$
$\mathbb{C}$ or $\mathbb{H}$, we call $\mathbf{V}$ a \emph{vector space} over
$\mathbb{K}$. When $\mathbb{K=H}$ it is necessary to distinguish between right
or left $\mathbb{H}$-linear spaces and in this case $\mathbf{V}$ will be
called a right or left $\mathbb{H}$-module. Recall that $\mathbb{H}$ is a
division ring (sometimes called a noncommutative field or a skew field) and
since $\mathbb{H}$ has a natural vector space structure over the real field,
then $\mathbb{H}$ is also a division algebra.

Let $\dim_{\mathbb{R}}\mathbf{V}=2m=n$. In this case it is possible to give the

\begin{definition}
A linear mapping
\end{definition}

\begin{equation}
\mathbf{J}:\mathbf{V\rightarrow V} \tag{B1}%
\end{equation}
\emph{such that}
\begin{equation}
\mathbf{J}^{2}=-\mathrm{I}\text{\textrm{d}}_{\mathbf{V}}, \tag{B2}%
\end{equation}
\emph{is called a complex structure mapping}.\medskip

\begin{definition}
The pair $\left(  \mathbf{V,J}\right)  $ will be called a complex vector space
structure and denote by $\mathbf{V}_{\mathbb{C}}$ if the following product
holds. Let $\mathbb{C}\ni z=a+ib$ and let $\mathbf{v\in V}$. Then
\end{definition}

\begin{equation}
z\mathbf{v}=(a+ib)\mathbf{v}=a\mathbf{v}+b\mathbf{Jv}. \tag{B3}%
\end{equation}
\medskip

It is obvious that $\dim_{\mathbb{C}}=\frac{m}{2}$.

\begin{definition}
Let $\mathbf{V}$ be a vector space over $\mathbb{R}$. A
\emph{complexification} of $\mathbf{V}$ is a complex structure associated with
the real vector space $\mathbf{V}\oplus\mathbf{V}$. The resulting complex
vector space is denoted by $\mathbf{V}^{\mathbb{C}}$. Let $\mathbf{v,w\in V}$.
Elements of $\mathbf{V}^{\mathbb{C}}$ are usually denoted by $\mathbf{c=v+}%
i\mathbf{w}$, and if $\mathbb{C}\ni z=a+ib$ we have
\end{definition}

\begin{equation}
z\mathbf{c}=a\mathbf{v}-b\mathbf{w}+i(a\mathbf{w}+b\mathbf{v}). \tag{B4}%
\end{equation}
\medskip

Of course, we have that $\dim_{\mathbb{C}}\mathbf{V}^{\mathbb{C}}%
=\dim_{\mathbb{R}}\mathbf{V}$.

\begin{definition}
A $\mathbb{H}$-module is a real vector space $\mathbf{V}$ carrying three
linear transformation, $\mathbf{I}$\textbf{, }$\mathbf{J}$ and $\mathbf{K}$
each one of them satisfying
\end{definition}

\begin{align}
\mathbf{I}^{2}  &  =\mathbf{J}^{2}=-\mathbf{I}\text{\textrm{d}}_{\mathbf{S}%
},\nonumber\\
\mathbf{IJ}  &  =-\mathbf{JI}=\mathbf{K},\text{ }\mathbf{JK}=-\mathbf{KJ}%
=\mathbf{I,}\text{ }\mathbf{KI}=-\mathbf{IK=J.} \tag{B5}%
\end{align}
\medskip

\begin{definition}
Any subset $I\subseteq\mathcal{A}$ such that
\end{definition}

\[
a\psi\in I,\forall a\in\mathcal{A},\text{ }\forall\psi\in I,
\]
\begin{equation}
\psi+\phi\in I,\forall\psi,\phi\in I \tag{B6}%
\end{equation}
\emph{is called a left ideal of }$\mathcal{A}$.

\begin{remark}
An analogous definition holds for \emph{right} ideals where Eq.(B6) reads
$\psi a\in I,\forall a\in\mathcal{A},$ $\forall\psi\in I$, for
\emph{bilateral} ideals where in this case Eq.(B6) reads $a\psi b\in I,\forall
a,b\in\mathcal{A},$ $\forall\psi\in I$.
\end{remark}

\begin{definition}
An associative $\mathcal{A}$ algebra on the \emph{field }$\mathbb{F}$
$\mathbb{(R}$ or $\mathbb{C)}$ is \emph{simple} if the only bilateral ideals
are the zero ideal and $\mathcal{A}$ itself.
\end{definition}

We give without proofs the following theorems.

\begin{theorem}
All minimal left (respectively right) ideals of $\mathcal{A}$ are of the form
$J=\mathcal{A}\mathrm{e}$ (respectively \textrm{e}$\mathcal{A}$), where
\textrm{e} is a primitive idempotent of $\mathcal{A}$.
\end{theorem}

\begin{theorem}
Two minimal left ideals of $\mathcal{A}$, $J=\mathcal{A}\mathrm{e}$ and
$J=\mathcal{A}\mathrm{e}^{\prime}$ are \emph{isomorphic} if and only if there
exist a non null $X^{\prime}\in J^{\prime}$ such that $J^{\prime}=JX^{\prime}
$.

We recall that \textrm{e}$\mathbf{\in}\mathcal{A}$ is an \emph{idempotent}
element if \textrm{e}$^{2}=\mathrm{e}$. An idempotent is said to be
\emph{primitive} if it cannot be written as the sum of two non zero
annihilating (or orthogonal) idempotent, i.e., $\mathrm{e}$ $\mathbf{\neq}$
$\mathrm{e}_{1}+\mathrm{e}_{2}$, with $\mathrm{e}_{1}\mathrm{e}_{2}%
=\mathrm{e}_{2}\mathrm{e}_{1}=0$ and $\mathrm{e}_{1}^{2}=\mathrm{e}_{1},$
$\mathrm{e}_{2}^{2}=\mathrm{e}_{2}$.
\end{theorem}

Not all algebras are simple and in particular \emph{semi-simple} algebras are
important for our considerations. A definition of semi simple algebras
requires the introduction of the concepts of \emph{nilpotent} ideals and
radicals. To define these concepts adequately would lead us to a long
incursion on the theory of associative algebras, so we avoid to do that here.
We only quote that semi-simple algebras are the direct sum of simple algebras.
Then, the study of semi-simple algebras is reduced to the study of simple algebras.

Now, let $\mathcal{A}$ be an associative and simple algebra on the
\emph{field}$\mathbb{\ F}$ $(\mathbb{R}$ or $\mathbb{C})$, and let
$\mathbf{S}$ be a finite dimensional linear space over a division ring
$\mathbb{K\subseteq F}$.

\begin{definition}
A \emph{representation} of $\mathcal{A}$ in $\mathbf{S}$ is a $\mathbb{K}$
algebra homomorphism\footnote{We recall that a $\mathbb{K}$-algebra
homomorphism is a $\mathbb{K}$-linear map $\rho$ such that $\forall
X,Y\in\mathcal{A},$ $\rho(XY)=\rho(X)\rho(Y)$.} $\rho:\mathcal{A}%
\rightarrow\mathbf{E=}\mathrm{End}_{\mathbb{K}}\mathbf{S}$ ($\mathbf{E=}%
\mathrm{End}_{\mathbb{K}}\mathbf{S}=\mathrm{Hom}_{\mathbb{K}}(\mathbf{S,S})$
is the endomorphism algebra of $\mathbf{S}$) which maps the unit element of
$\mathcal{A}$ to $\mathbf{I}\mathrm{d}_{\mathbf{E}}$. the dimension
$\mathbb{K}$ of $\mathbf{S}$ is called the \emph{degree} of the representation.
\end{definition}

The addition in $\mathbf{S}$ together with the mapping $\mathcal{A}%
\times\mathbf{S\rightarrow S}$, $(a,x)\mapsto\rho(a)x$ turns $\mathbf{S}%
$\textbf{\ }in a left $\mathcal{A}$-module\footnote{We recall that there are
left and right modules, so we can also define right modular representations of
$\mathcal{A}$ by defining the mapping $\mathbf{S}\times\mathcal{A}%
\mathbf{\rightarrow S}$, $(x,a)\mapsto x\rho(a)$. This turns $\mathbf{S}%
$\textbf{\ }in a right $\mathcal{A}$-module, called the right
\emph{representation module}.}, called the left \emph{representation module}.

\begin{remark}
It is important to recall that when $\mathbb{K=H}$ the usual recipe for
$\mathrm{Hom}_{\mathbb{H}}(\mathbf{S,S})$ to be a linear space over
$\mathbb{H}$ fails and in general $\mathrm{Hom}_{\mathbb{H}}(\mathbf{S,S})$ is
considered as a linear space over $\mathbb{R}$, which is the centre of
$\mathbb{H}$.
\end{remark}

\begin{remark}
We also have that if $\mathcal{A}$ is an algebra over $\mathbb{F}$ and
$\mathbf{S}$ is an $\mathcal{A}$-module, then $\mathbf{S}$ can always be
considered as a vector space over $\mathbb{F}$ and if $e\in\mathcal{A}$, the
mapping $\chi:a\rightarrow\chi_{a}$ with $\chi_{a}(\mathbf{s})=a\mathbf{s},$
$\mathbf{s\in S}$, is a homomorphism $\mathcal{A}\rightarrow\mathbf{E=}%
\mathrm{End}_{\mathbb{F}}\mathbf{S}$, and so it is a representation of
$\mathcal{A}$ in $\mathbf{S}$. The study of $\mathcal{A}$ modules is then
equivalent to the study of the $\mathbb{F}$ representations of $\mathcal{A}$.
\end{remark}

\begin{definition}
A representation $\rho$ is \emph{faithful} if its kernel is zero, i.e.,
$\rho(a)x=0,\forall x\in\mathbf{S}\Rightarrow a=0$. The kernel of $\rho$ is
also known as the \emph{annihilator} of its module.
\end{definition}

\begin{definition}
$\rho$ is said to be \emph{simple} or irreducible if the only invariant
subspaces of $\rho(a),$ $\forall a\in\mathcal{A}$, are $\mathbf{S}$ and
$\left\{  0\right\}  $.
\end{definition}

Then, the representation module is also simple. That means that it has no
proper submodules.

\begin{definition}
$\rho$ is said to be \emph{semi-simple}, if it is the direct sum of simple
modules, and in this case $\mathbf{S}$ is the direct sum of subspaces which
are globally invariant under $\rho(a),$ $\forall a\in\mathcal{A}$.
\end{definition}

When no confusion arises $\rho(a)x$ may be denoted by $a\bullet x,$ $a*x$ or
$ax$.

\begin{definition}
Two $\mathcal{A}$-modules $\mathbf{S}$ and $\mathbf{S}^{\prime}$ (with the
exterior multiplication being denoted respectively by $\bullet$ and $*$) are
\emph{isomorphic }if there exists a bijection $\varphi:\mathbf{S\rightarrow
S}^{\prime}$ such that,
\end{definition}

\begin{align*}
\varphi(x+y)  &  =\varphi(x)+\varphi(y),\text{ }\forall x,y\in\mathbf{S,}\\
\varphi(a\bullet x)  &  =a*\varphi(x),\text{ }\forall a\in\mathcal{A},
\end{align*}
\emph{and we say that representation }$\rho$\emph{\ and }$\rho^{\prime}%
$\emph{\ of }$\mathcal{A}$\emph{\ are equivalent if their modules are
isomorphic.\medskip}

This implies the existence of a $\mathbb{K}$-linear isomorphism $\varphi
:\mathbf{S\rightarrow S}^{\prime}$ such that $\varphi\circ\rho(a)=\rho
^{\prime}(a)\circ\varphi,$ $\forall a\in\mathcal{A}$ or $\rho^{\prime
}(a)=\varphi\circ\rho(a)\circ\varphi^{-1}$. If $\dim\mathbf{S}=n$, then
$\dim\mathbf{S}^{\prime}=n$.

\begin{definition}
A \emph{complex representation} of $\mathcal{A}$ is simply a real
representation $\rho:\mathcal{A}\rightarrow\mathrm{Hom}_{\mathbb{R}%
}(\mathbf{S,S})$ for which
\end{definition}

\begin{equation}
\rho(X)\circ\mathbf{J=J}\circ\rho(X),\text{ }\forall X\in\mathcal{A}. \tag{B7}%
\end{equation}
\medskip

This means that the image of $\rho$ commutes with the subalgebra generated by
$\left\{  \mathbf{I}\mathrm{d}_{\mathbf{S}}\right\}  \sim\mathbb{C}$.

\begin{definition}
A quaternionic representation of $\mathcal{A}$ is a representation
$\rho:\mathcal{A}\rightarrow\mathrm{Hom}_{\mathbb{R}}(\mathbf{S,S})$ such that
\end{definition}

\begin{equation}
\rho(X)\circ\mathbf{I=I}\circ\rho(X),\text{ }\rho(X)\circ\mathbf{J=J\circ}%
\rho(X),\text{ }\rho(X)\circ\mathbf{K=K\circ}\rho(X),\forall X\in\mathcal{A}.
\tag{B8}%
\end{equation}
\medskip This means that the representation $\rho$ has a commuting subalgebra
isomorphic to the quaternion ring.

The following theorem (\cite{16},\cite{39}) is crucial:

\begin{theorem}
\textbf{(Wedderburn)} If $\mathcal{A}$ is simple algebra over $\mathbb{F}$
then $\mathcal{A}$ is isomorphic to $\mathbb{D}(m)$, where $\mathbb{D}(m)$ is
a matrix algebra with entries in $\mathbb{D}$ (a division algebra), and $m $
and $\mathbb{D}$ are unique (modulo isomorphisms).
\end{theorem}

Now, it is time to specialize our results to the Clifford algebras over the
field $\mathbb{F=R}$ or $\mathbb{C}$. We are particularly interested in the
case of real Clifford algebras. In what follows we take $(\mathbf{V}%
,\mathbf{b})=(\mathbb{R}^{n},\mathbf{g})$. We denote by $\mathbb{R}^{p,q}$ a
real vector space of dimension $n=p+q$ endowed with a nondegenerate metric
$\mathbf{g}:\mathbb{R}^{n}\times\mathbb{R}^{n}\rightarrow\mathbb{R}$. Let
$\left\{  E_{i}\right\}  ,$ $(i=1,2,...,n)$ be an orthonormal basis of
$\mathbb{R}^{p,q}$,
\begin{equation}
\mathbf{g}(E_{i},E_{j})=g_{ij}=g_{ji}=\left\{
\begin{array}
[c]{l}%
+1,\text{ }i=j=1,2,...p\\
-1,\text{ }i=j=p+1,...,p+q=n\\
\text{ \thinspace\thinspace}0,\text{ }i\neq j
\end{array}
\right.  \tag{B9}%
\end{equation}

\begin{definition}
The Clifford algebra $\mathbb{R}_{p,q}=\mathcal{C}\ell(\mathbb{R}^{p,q})$ is
the Clifford algebra over $\mathbb{R}$, generated by 1 and the $\left\{
E_{i}\right\}  ,$ $(i=1,2,...,n)$ such that $E_{i}^{2}=\mathbf{q}%
(E_{i})=\mathbf{g}(E_{i},E_{i})$, $E_{i}E_{j}=-E_{j}E_{i}$ $(i\neq j)$, and
$E_{1}E_{2}...E_{n}\neq\pm1$.
\end{definition}

$\mathbb{R}_{p,q}$ is obviously of dimension $2^{n}$ and as a vector space it
is the direct sum of vector spaces $\bigwedge^{k}\mathbb{R}^{n}$ of dimensions
$\binom{n}{k},0\leq k\leq n$. The canonical basis of $\bigwedge^{k}%
\mathbb{R}^{n}$ is given by the elements $e_{A}=E_{\alpha_{1}}...E_{\alpha
_{k}},$ $1\leq\alpha_{1}<...<\alpha_{k}\leq n$. The element $e_{J}%
=E_{1}...E_{n}\in\bigwedge^{k}\mathbb{R}^{n}\subset\mathbb{R}_{p,q}$ commutes
($n$ odd) or anticommutes ($n$ even) with all vectors $E_{1}...E_{n}%
\in\bigwedge^{1}\mathbb{R}^{n}\equiv\mathbb{R}^{n}$. The center $\mathcal{C}%
\ell_{p,q}$ is $\bigwedge^{0}\mathbb{R}^{n}\equiv\mathbb{R}$ if $n$ is even
and it is the direct sum $\bigwedge^{0}\mathbb{R}^{n}\oplus\bigwedge
^{0}\mathbb{R}^{n}$ if $n$ is odd.

All Clifford algebras are semi-simple. If $p+q=n$ is even, $\mathbb{R}_{p,q}$
is simple and if $p+q=n$ is odd we have the following possibilities:

(a) $\mathbb{R}_{p,q}$ is simple $\leftrightarrow c_{J}^{2}=-1\leftrightarrow
p-q\neq1$ (mod 4) $\leftrightarrow$ center of $\mathbb{R}_{p,q}$ is isomorphic
to $\mathbb{C}$;

(b) $\mathbb{R}_{p,q}$ is not simple (but is a direct sum of two simple
algebras) $\leftrightarrow c_{J}^{2}=+1\leftrightarrow p-q=1$ (mod 4)
$\leftrightarrow$ center of $\mathbb{R}_{p,q}$ is isomorphic to
$\mathbb{R\oplus R} $.

Now, for $\mathbb{R}_{p,q}$ the division algebras $\mathbb{D}$ are the
division rings $\mathbb{R}$, $\mathbb{C}$ or $\mathbb{H}$. The explicit
isomorphism can be discovered with some hard but not difficult work. It is
possible to give a general classification off all real (and also the complex)
Clifford algebras and a classification table can be found, e.g., in
(\cite{45},\cite{46}). Such a table is reproduced below and $\left[  \frac
{n}{2}\right]  $ means the integer part of $n/2$.

\bigskip

\hspace{-3.2cm}%
\[
\hspace{-1.2cm}\hspace{-0.5cm}%
\begin{array}
[c]{lllllllll}%
\begin{array}
[c]{l}%
_{p-q}\\
\text{mod 8}%
\end{array}
& \hspace{0.4cm}0 & \hspace{0.4cm}1 & \hspace{0.4cm}2 & \hspace{0.4cm}3 &
\hspace{0.4cm}4 & \hspace{0.4cm}5 & \hspace{0.4cm}6 & \hspace{0.4cm}7\\
\mathbb{R}_{p,q} & _{\mathbb{R}(2^{\left[  \frac{n}{2}\right]  })} &
\begin{array}
[c]{l}%
_{\mathbb{R}(2^{\left[  \frac{n}{2}\right]  })}\\
\hspace{0.4cm}\oplus\\
_{\mathbb{R}(2^{\left[  \frac{n}{2}\right]  })}%
\end{array}
& _{\mathbb{R}(2^{\left[  \frac{n}{2}\right]  })} & _{\mathbb{C}(2^{\left[
\frac{n}{2}\right]  })} & _{\mathbb{H}(2^{\left[  \frac{n}{2}\right]  -1})} &
\begin{array}
[c]{l}%
_{\mathbb{H}(2^{\left[  \frac{n}{2}\right]  -1})}\\
\hspace{0.4cm}\oplus\\
_{\mathbb{H}(2^{\left[  \frac{n}{2}\right]  -1})}%
\end{array}
& _{\mathbb{H}(2^{\left[  \frac{n}{2}\right]  -1})} & _{\mathbb{C}(2^{\left[
\frac{n}{2}\right]  })}%
\end{array}
\]

\begin{center}
\textbf{Table 1.} Representation of the Clifford algebras $\mathbb{R}_{p,q}$
as matrix algebras
\end{center}

Now, to complete the classification we need the following theorem\cite{45}.

\begin{theorem}
\textbf{\ (Periodicity) }
\end{theorem}

\begin{equation}%
\begin{array}
[c]{lllllll}%
\mathbb{R}_{n+8} & = & \mathbb{R}_{n,0}\otimes\mathbb{R}_{8,0} &  &
\mathbb{R}_{0,n+8} & = & \mathbb{R}_{0,n}\otimes\mathbb{R}_{0,8}\\
\mathbb{R}_{p+8,q} & = & \mathbb{R}_{p,q}\otimes\mathbb{R}_{8,0} &  &
\mathbb{R}_{p,q+8} & = & \mathbb{R}_{p,q}\otimes\mathbb{R}_{0,8}%
\end{array}
\tag{B10}%
\end{equation}
\medskip

\begin{remark}
We emphasize here that since the general results concerning the
representations of simple algebras over a field $\mathbb{F}$ applies to the
Clifford algebras $\mathbb{R}_{p,q}$ we can talk about real, complex or
quaternionic representation of a given Clifford algebra, even if the natural
matrix identification is not a matrix algebra over one of these fields. A case
that we shall need is that $\mathbb{R}_{1,3}\simeq\mathbb{H}(2)$. But it is
clear that $\mathbb{R}_{1,3}$ has a complex representation, for any
quaternionic representation of $\mathbb{R}_{p,q}$ is automatically
\emph{complex}, once we restrict $\mathbb{C\subset H}$ and of course, the
complex dimension of any $\mathbb{H}$-module must be even. Also, any complex
representation of $\mathbb{R}_{p,q}$ extends automatically to a representation
of $\mathbb{C\otimes R}_{p,q}$.
\end{remark}

\begin{remark}
\label{a}Now, $\mathbb{C\otimes R}_{p,q}$ is an abbreviation for the
\emph{complex} Clifford algebra $\mathcal{C}\ell_{p+q}=\mathbb{C\otimes
R}_{p,q}$, i.e., it is the tensor product of the algebras $\mathbb{C}$ and
$\mathbb{R}_{p,q}$, which are subalgebras of the finite dimensional algebra
$\mathcal{C}\ell_{p+q}$ over $\mathbb{C}$.
\end{remark}

For the purposes of the present paper we shall need to have in mind that
\begin{align}
\mathbb{R}_{0,1}  &  \simeq\mathbb{C}\nonumber\\
\mathbb{R}_{0,2}  &  \simeq\mathbb{H}\nonumber\\
\mathbb{R}_{3,0}  &  \simeq\mathbb{C}(2)\tag{B11}\\
\mathbb{R}_{1,3}  &  \simeq\mathbb{H}(2)\nonumber\\
\mathbb{R}_{3,1}  &  \simeq\mathbb{R}(4)\nonumber\\
\mathbb{R}_{4,1}  &  \simeq\mathbb{C}(4)\nonumber
\end{align}

$\mathbb{R}_{3,0}$ is called the Pauli algebra, $\mathbb{R}_{1,3}$ is called
the \emph{spacetime} algebra, $\mathbb{R}_{3,1}$ is called \emph{Majorana}
algebra and $\mathbb{R}_{4,1}$ is called the \emph{Dirac} algebra. Also the
following particular results have been used in the text and below.
\begin{align}
\mathbb{R}_{1,3}^{0}  &  \simeq\mathbb{R}_{3,1}^{0}=\mathbb{R}_{3,0},\text{
}\mathbb{R}_{4,1}^{0}\simeq\mathbb{R}_{1,3},\nonumber\\
\mathbb{R}_{4,1}  &  \simeq\mathbb{C\otimes R}_{3,1}\text{ }\mathbb{R}%
_{4,1}\simeq\mathbb{C\otimes R}_{3,1}, \tag{B12}%
\end{align}
which means that the Dirac algebra is the complexification of both the
spacetime or the Majorana algebras.

Eq.(B11) show moreover, in view of Remark 7 that the spacetime algebra has a
complexification matrix representation in $\mathbb{C}(4)$. Obtaining such a
representation is fundamental for the present work and it is given in Appendix D.

\subsection{Minimal Lateral Ideals of $\mathbb{R}_{p,q}$}

It is important for the objectives of this paper to know some results
concerning the minimal lateral ideals of $\mathbb{R}_{p,q}$. The
identification table of these algebras as matrix algebras helps a lot. Indeed,
we have \cite{16}

\begin{theorem}
The maximum number of pairwise orthogonal idempotents in $\mathbb{K}(m)$
(where $\mathbb{K=R}$, $\mathbb{C}$ or $\mathbb{H}$) is $m$ .
\end{theorem}

The decomposition of $\mathbb{R}_{p,q}$ into minimal ideals is then
characterized by a spectral set \{$\mathrm{e}_{pq,j}$\} of idempotents
elements of $\mathbb{R}_{p,q}$ such that

(a) $\sum\limits_{i=1}^{n}\mathrm{e}_{pq,i}=1$;

(b)$\mathrm{e}_{pq,j}\mathrm{e}_{pq,k}=\delta_{jk}\mathrm{e}_{pq,j}$;

(c) the rank of $\mathrm{e}_{pq,j}$ is minimal and non zero, i.e., is primitive.

By rank of $\mathrm{e}_{pq,j}$ we mean the rank of the $\bigwedge
\mathbb{R}^{p,q}$ morphism, $\mathrm{e}_{pq,j}:\phi\longmapsto\phi
\mathrm{e}_{pq,j}$. Conversely, any $\phi\in\mathbf{I}_{pq,j}$ can be
characterized by an idempotent\textrm{\ }$\mathrm{e}_{pq,j}$ of minimal rank
$\neq0$, with $\phi=\phi\mathrm{e}_{pq,j}$.

We now need to know the following theorem \cite{39}.

\begin{theorem}
A minimal left ideal of $\mathbb{R}_{p,q}$ is of the type
\end{theorem}

\begin{equation}
\mathbf{I}_{pq}=\mathbb{R}_{p,q}\mathrm{e}_{pq} \tag{B12}%
\end{equation}
\textit{where}%

\begin{equation}
\mathrm{e}_{pq}=\frac{1}{2}(1+e_{\alpha_{1}})...\frac{1}{2}(1+e_{\alpha_{k}})
\tag{B13}%
\end{equation}
\textit{is a primitive idempotent of }$\mathbb{R}_{p,q}$\textit{\ and were
}$e_{\alpha_{1}},...,e_{\alpha_{k}}$\textit{\ are commuting elements in the
canonical basis of }$\mathbb{R}_{p,q}$\textit{\ generated in the standard way
through the elements of the basis }$\sum$\textit{\ such that }$(e_{\alpha_{i}%
})^{2}=1,$\textit{\ }$(i=1,2,...,k)$\textit{\ generate a group of order
}$2^{k}$\textit{, }$k=q-r_{q-p}$\textit{\ and }$r_{i}$\textit{\ are the
Radon-Hurwitz numbers, defined by the recurrence formula }$r_{i+8}=r_{i}%
+4$\textit{\ and }%

\begin{equation}%
\begin{tabular}
[c]{|c|c|c|c|c|c|c|c|c|c|}\hline
$i$ &  & $0$ & $1$ & $2$ & $3$ & $4$ & $5$ & $6$ & $7$\\\hline
$r_{i}$ &  & $0$ & $1$ & $2$ & $2$ & $3$ & $3$ & $3$ & $3$\\\hline
\end{tabular}
. \tag{B14}%
\end{equation}
\medskip

Recall that $\mathbb{R}_{p,q}$ is a ring and the minimal lateral ideals are
modules over the ring $\mathbb{R}_{p,q}$. They are \emph{representation
modules} of $\mathbb{R}_{p,q}$, and indeed we have (recall the table above)
the theorem \cite{45}

\begin{theorem}
If $p+q$ is even or odd with $p-q\neq1$ (mod 4), then
\end{theorem}

\begin{equation}
\mathbb{R}_{p,q}=\mathrm{Hom}_{\mathbb{K}}(I_{pq})\simeq\mathbb{K}(m),
\tag{B15}%
\end{equation}
\emph{where (as we already know) }$\mathbb{K=R}$\emph{, }$\mathbb{C}%
$\emph{\ or }$\mathbb{H}$\emph{. Also},
\begin{equation}
\mathrm{\dim}_{\mathbb{K}}(I_{pq})=m, \tag{B16}%
\end{equation}
\emph{and}
\begin{equation}
\mathbb{K}\simeq\mathrm{e}\mathbb{K}(m)\mathrm{e,} \tag{B17}%
\end{equation}
\emph{where }e\emph{\ is the representation of }$\mathbf{e}_{pq}$\emph{\ in
}$\mathbb{K}(m)$.

\emph{If }$p+q=n$\emph{\ is odd, with }$p-q=1$\emph{\ (mod 4), then}
\begin{equation}
\mathbb{R}_{p,q}=\mathrm{Hom}_{\mathbb{K}}(I_{pq})\simeq\mathbb{K}%
(m)\oplus\mathbb{K}(m), \tag{B18}%
\end{equation}
\emph{with}
\begin{equation}
\mathrm{\dim}_{\mathbb{K}}(I_{pq})=m \tag{B19}%
\end{equation}
\emph{and}
\begin{align}
\mathrm{e}\mathbb{K}(m)\mathrm{e}  &  \simeq\mathbb{R\oplus R}\nonumber\\
&  \text{or}\tag{B20}\\
\mathrm{e}\mathbb{K}(m)\mathrm{e}  &  \simeq\mathbb{H\oplus H}.\nonumber
\end{align}
\medskip With the above isomorphisms we can immediately identify the minimal
left ideals of $\mathbb{R}_{p,q}$ with the column matrices of $\mathbb{K}(m)$.

\subsubsection{Algorithm for Finding Primitive Idempotents of $\mathbb{R}%
_{p,q}$.}

With the ideas introduced above it is now a simple exercise to find primitive
idempotents of $\mathbb{R}_{p,q}$. First we look at \textbf{Table 1} and find
the matrix algebra to which our particular Clifford algebra $\mathbb{R}_{p,q}$
is isomorphic. Suppose $\mathbb{R}_{p,q}$ is simple\footnote{Once we know the
algorithm for a simple Clifford algebra it is straightfoward to devise an
algorithm for the semi-simple Clifford algebras.}. Let $\mathbb{R}_{p,q}%
\simeq$ $\mathbb{K}(m)$ for a particular $\mathbb{K}$ and $m$. Next we take an
element $e_{\alpha_{1}}\in\left\{  e_{A}\right\}  $ from the canonical basis
$\left\{  e_{A}\right\}  $ of $\mathbb{R}_{p,q}$ such that
\begin{equation}
e_{\alpha_{1}}^{2}=1. \tag{B21}%
\end{equation}
then construct the idempotent $\mathrm{e}_{pq}=(1+e_{\alpha_{1}})/2$ and the
ideal $\mathbf{I}_{pq}=\mathbb{R}_{p,q}\mathrm{e}_{pq}$ and calculate
$\mathrm{\dim}_{\mathbb{K}}(I_{pq})$. If $\mathrm{\dim}_{\mathbb{K}}(I_{pq})=m
$, then $\mathrm{e}_{pq}$ is primitive. If $\mathrm{\dim}_{\mathbb{K}}%
(I_{pq})\neq m$, we choose $e_{\alpha2}\in\left\{  e_{A}\right\}  $ such that
$e_{\alpha2}$ commutes with $e_{\alpha_{1}}$ and $e_{\alpha_{2}}^{2}=1$ (see
Theorem 39 and construct the idempotent $\mathrm{e}_{pq}^{\prime}%
=(1+e_{\alpha_{1}})(1+e_{\alpha_{1}})/4$. If $\mathrm{\dim}_{\mathbb{K}%
}(I_{pq}^{\prime})=m$, then $\mathrm{e}_{pq}^{\prime}$ is primitive. Otherwise
we repeat the procedure. According to the Theorem 39 the procedure is finite.

These results will be used in Appendix D in order to obtain necessary results
for our presentation of the theory of algebraic and Dirac-Hestenes spinors
(and spinors fields).

\section{$\mathbb{R}_{p,q}^{\star}$, Clifford, Pinor and Spinor Groups}

The set of the invertible elements of $\mathbb{R}_{p,q}$ constitutes a
non-abelian group which we denote by $\mathbb{R}_{p,q}^{\star}$. It acts
naturally on $\mathbb{R}_{p,q}$ as an algebra homomorphism through its adjoint
representation
\begin{equation}
\mathrm{Ad}:\mathbb{R}_{p,q}^{\star}\rightarrow\mathrm{Aut}(\mathbb{R}%
_{p,q});\text{ }u\longmapsto\mathrm{Ad}_{u},\text{ with }\mathrm{Ad}%
_{u}(x)=uxu^{-1}. \tag{C1}%
\end{equation}
We define also the twisted adjoint representation:%
\begin{equation}
\mathrm{\hat{A}d}:\mathbb{R}_{p,q}^{\star}\rightarrow\mathrm{Aut}%
(\mathbb{R}_{p,q});\text{ }u\longmapsto\mathrm{\hat{A}d},\text{ with
}\mathrm{\hat{A}d}_{u}(x)=ux\hat{u}^{-1} \tag{C1'}%
\end{equation}

\begin{definition}
The \emph{Clifford-Lipschitz} group is the set
\end{definition}

\begin{equation}
\Gamma_{p,q}=\left\{  u\in\mathbb{R}_{p,q}^{\star}\left\vert \forall
x\in\mathbb{R}^{p,q},\text{ }ux\hat{u}^{-1}\in\mathbb{R}^{p,q}\right.
\right\}  . \tag{C2}%
\end{equation}
\medskip

\begin{definition}
The set $\Gamma_{p,q}^{0}=\Gamma_{p,q}\cap\mathbb{R}_{p,q}^{0}$ is called
\emph{special} Clifford-Lipshitz group.
\end{definition}

\begin{definition}
The \emph{Pinor group} \textrm{Pin}$_{p.q}$ is the subgroup of $\ \Gamma
_{p,q}$ such that
\end{definition}

\begin{equation}
\mathrm{Pin}_{p,q}=\left\{  u\in\Gamma_{p,q}\text{
%TCIMACRO{\TEXTsymbol{\vert} }%
%BeginExpansion
$\vert$
%EndExpansion
}N(u)=\pm1\right\}  , \tag{C3}%
\end{equation}
\medskip%
\[
N:\mathbb{R}_{p,q}\rightarrow\mathbb{R}_{p,q},\text{ }N(x)=\langle\bar
{x}x\rangle_{0}%
\]

\begin{definition}
The \emph{Spin group} $\mathrm{Spin}_{p,q}$ is the set
\end{definition}

\begin{equation}
\mathrm{Spin}_{p,q}=\left\{  u\in\Gamma_{p,q}^{0}\text{
%TCIMACRO{\TEXTsymbol{\vert} }%
%BeginExpansion
$\vert$
%EndExpansion
}N(u)=\pm1\right\}  . \tag{C4}%
\end{equation}
\medskip

It is easy to see that $\mathrm{Spin}_{p,q}$ is not connected.

\begin{definition}
The group $\mathrm{Spin}_{p,q}^{e}$ is the set
\end{definition}

\begin{equation}
\mathrm{Spin}_{p,q}^{e}=\left\{  u\in\Gamma_{p,q}^{0}\text{
%TCIMACRO{\TEXTsymbol{\vert} }%
%BeginExpansion
$\vert$
%EndExpansion
}N(u)=+1\right\}  . \tag{C5}%
\end{equation}
\medskip

The superscript $e$, means that $\mathrm{Spin}_{p,q}^{e}$ is the connected
component to the identity. We can prove that $\mathrm{Spin}_{p,q}^{e}$ is
connected for all pairs $(p,q)$ with the exception of $\mathrm{Spin}%
^{e}(1,0)\simeq\mathrm{Spin}^{e}(0,1)$.

We recall now some classical results \cite{40} associated with the
pseudo-orthogonal groups \textrm{O}$_{p,q}$ of a vector space $\mathbb{R}%
^{p,q}$ ($n=p+q$) and its subgroups.

Let $\mathbf{G}$ be a diagonal $n\times n$ matrix whose elements are
\begin{equation}
G_{ij}=\mathrm{diag}(1,1,...,-1,-1,...-1), \tag{C6}%
\end{equation}

with $p$ positive and $q$ negative numbers.

\begin{definition}
\textrm{O}$_{p,q}$ is the set of $n\times n$ real matrices $\mathbf{L}$ such that
\end{definition}

\begin{equation}
\mathbf{LGL}^{T}=\mathbf{G},\text{ }\det\mathbf{L}^{2}=1. \tag{C7}%
\end{equation}
$\medskip$

Eq.(C7) shows that \textrm{O}$_{p,q}$ is not connected.

\begin{definition}
\textrm{SO}$_{p,q}$, the special (proper) pseudo orthogonal group is the set
of $n\times n$ real matrices $\mathbf{L}$ such that
\end{definition}

\begin{equation}
\mathbf{LGL}^{T}=\mathbf{G},\text{ }\det\mathbf{L}=1. \tag{C8}%
\end{equation}
\medskip

When $p=0$ ($q=0$) \textrm{SO}$_{p,q}$ is connected. However, \textrm{SO}%
$_{p,q}$ is not connected and has two connected components for $p,q\geq1$. The
group \textrm{SO}$_{p,q}^{e}$, the connected component to the identity of
\textrm{SO}$_{p,q}$ will be called the special \emph{orthocronous}
pseudo-orthogonal group\footnote{This nomenclature comes from the fact that
SO$^{e}(1,3)=\mathcal{L}_{+}^{\uparrow}$ is the special (proper) orthochronous
Lorentz group. In this case the set is easily defined by the condition
$L_{0}^{0}\geq+1$. For the general case see \cite{40} .}.\medskip

\begin{theorem}
\textbf{:} $\mathrm{\hat{A}d}_{\left\vert \mathrm{Pin}_{p,q}\right.
}:\mathrm{Pin}_{p,q}\rightarrow\mathrm{O}_{p,q}$ is onto with kernel
$\mathbb{Z}_{2}$. \textrm{Ad}$_{\left\vert \mathrm{Spin}_{p,q}\right.
}:\mathrm{Spin}_{p,q}\rightarrow\mathrm{SO}_{p,q}$ is onto with kernel
$\mathbb{Z}_{2}$. \textrm{Ad}$_{\left\vert \mathrm{Spin}_{p,q}^{e}\right.
}:\mathrm{Spin}_{p,q}^{e}\rightarrow\mathrm{SO}_{p,q}^{e}$ is onto with kernel
$\mathbb{Z}_{2}$. We have,
\begin{equation}
\mathrm{O}_{p,q}=\frac{\mathrm{Pin}_{p,q}}{\mathbb{Z}_{2}},\text{ }%
\mathrm{SO}_{p,q}=\frac{\mathrm{Spin}_{p,q}}{\mathbb{Z}_{2}},\text{
}\mathrm{SO}_{p,q}^{e}=\frac{\mathrm{Spin}_{p,q}^{e}}{\mathbb{Z}_{2}}.
\tag{C9}%
\end{equation}
\medskip
\end{theorem}

The group homomorphism between $\mathrm{Spin}_{p,q}^{e}$ and $\mathrm{SO}%
^{e}(p,q)$ will be denoted by
\begin{equation}
\mathbf{L}:\mathrm{Spin}_{p,q}^{e}\rightarrow\mathrm{SO}_{p,q}^{e}. \tag{C10}%
\end{equation}

The following theorem that first appears in Porteous book \cite{45} is very
important\footnote{In particular, when Theorem 49 is taken into account
together with some of the coincidence between the complexifications of some
low dimensions Clifford algebras (see Appendix C) it becomes clear that the
construction of Dirac-Hestenes spinors (and its representation as in eq.(D20))
for Minkowski vector space has no generalization for vector spaces of
arbitrary dimensions and signatures \cite{39}.}.

\begin{theorem}
(Porteous) For $p+q\leq5$, $\mathrm{Spin}^{e}(p,q)=\{u\in\mathbb{R}_{p,q}$
%TCIMACRO{\TEXTsymbol{\vert} }%
%BeginExpansion
$\vert$
%EndExpansion
$u\tilde{u}=1\}$.
\end{theorem}

\subsection{\textbf{Lie Algebra of }\textrm{Spin}$_{1,3}^{e}$}

It can be shown \cite{39} that for each $u\in\mathrm{Spin}_{1,3}^{e}$ it holds
$u=\pm e^{F},$ $F\in\bigwedge^{2}\mathbb{R}^{1,3}\subset\mathbb{R}_{1,3}$ and
$F$ can be chosen in such a way to have a positive sign in Eq.(C8), except in
the particular case $F^{2}=0$ when $u=-e^{F}$. From Eq.(C8) it follows
immediately that the Lie algebra of $\mathrm{Spin}_{1,3}^{e}$ is generated by
the bivectors $F\in\bigwedge^{2}\mathbb{R}^{1,3}\subset\mathbb{R}_{1,3}$
through the commutator product. More details on the realtions of Clifford
algebras and the rotation groups may be found, e.g., in \cite{ashdown,zeni}

\section{Spinor Representations of $\mathbb{R}_{4,1},$ $\mathbb{R}_{4,1}^{+}$
and $\mathbb{R}_{1,3}$}

Let $b_{0}=\left\{  E_{0},E_{1},E_{2},E_{3}\right\}  $ be an orthogonal
\ basis of $\mathbb{R}^{1,3}\subset\mathbb{R}_{1,3}$, such that $E_{\mu
}E_{\upsilon}+E_{\upsilon}E_{\mu}=2\eta_{\mu\nu}$, with $\eta_{\upsilon\mu
}=\mathrm{diag}(+1,-1,-1,-1)$. Now, with the results os Appendix B we can
verify without difficulties that the elements \textrm{e, }$\mathrm{e}^{\prime
},$ $\mathrm{e}^{\prime\prime}\in\mathbb{R}_{1,3}$%
\begin{align}
\mathrm{e}  &  =\frac{1}{2}(1+E_{0})\tag{D1}\\
\mathrm{e}^{\prime}  &  =\frac{1}{2}(1+E_{3}E_{0})\tag{D2}\\
\mathrm{e}^{\prime\prime}  &  =\frac{1}{2}(1+E_{1}E_{2}E_{3}) \tag{D3}%
\end{align}
are primitive idempotents of $\mathbb{R}_{1,3}$ The minimal left ideals,
$I=\mathbb{R}_{1,3}e$, $I^{\prime}=\mathbb{R}_{1,3}e^{\prime}$, $I^{\prime
\prime}=\mathbb{R}_{1,3}e^{\prime\prime}$ are \emph{right }two dimension
linear spaces over the quaternion field (e.g., $\mathbb{H}\mathrm{e}%
=\mathrm{e}\mathbb{H}=e\mathbb{R}_{1,3}\mathrm{e}$). According to a definition
given originally in \cite{49} these ideals are algebraically equivalent. For
example, $\mathrm{e}^{\prime}=u\mathrm{e}u^{-1}$, with $u=(1+E_{3}%
)\notin\Gamma_{1,3}$.

\begin{definition}
The elements $\Phi\in\mathbb{R}_{1,3}\frac{1}{2}(1+E_{0})$ are called
\emph{mother} spinors .
\end{definition}

The above denomination has been given (with justice) by Lounesto \cite{39}. It
can be shown (\cite{18},\cite{19}) that each $\Phi$ can be written
\begin{equation}
\Phi=\psi_{1}\mathrm{e+}\psi_{2}E_{3}E_{1}\mathrm{e+}\psi_{3}E_{3}%
E_{0}\mathrm{e+}\psi_{4}E_{1}E_{0}\mathrm{e=}\sum\limits_{i}\psi_{i}s_{i},
\tag{D4}%
\end{equation}%
\begin{equation}
s_{1}=\text{ }\mathrm{e,}\text{ }s_{2}=E_{3}E_{1}\mathrm{e,}\text{ }%
s_{3}=E_{3}E_{0}\mathrm{e,}\text{ }s_{4}=E_{1}E_{0}\mathrm{e} \tag{D5}%
\end{equation}
and where the $\psi_{i}$ are \emph{formally} complex numbers, i.e., each
$\psi_{i}=(a_{i}+b_{i}E_{2}E_{1})$ with $a_{i},$ $b_{i}\in\mathbb{R}$ and the
set $\left\{  s_{i},i=1,2,3,4\right\}  $ is a basis in the mother spinors space.

We recall from the general result of Appendix C that $\frac{\mathrm{Pin}%
_{1,3}}{\mathbb{Z}_{2}}\simeq\mathrm{O}_{1,3}$, $\frac{\mathrm{Spin}_{1,3}%
}{\mathbb{Z}_{2}}\simeq\mathrm{SO}_{1,3}$, $\frac{\mathrm{Spin}_{1,3}^{e}%
}{\mathbb{Z}_{2}}\simeq\mathrm{SO}_{1,3}^{e}$, and $\mathrm{Spin}_{1,3}%
^{e}\simeq\mathrm{Sl}(2,\mathbb{C})$ is the universal covering group of
$\mathcal{L}_{+}^{\uparrow}\equiv\mathrm{SO}_{1,3}^{e}$, the \emph{special}
(proper) \emph{orthocronous }Lorentz group.

In order to determine the relation between $\mathbb{R}_{4,1}$ and
$\mathbb{R}_{3,1} $ we proceed as follows: let $\left\{  F_{0},F_{1}%
,F_{2},F_{3},F_{4}\right\}  $ be an orthonormal basis of $\mathbb{R}_{4,1}$
with
\[
-F_{0}^{2}=F_{1}^{2}=F_{2}^{2}=F_{3}^{2}=F_{4}^{2}=1,F_{A}F_{B}=-F_{B}%
F_{A}(A\neq B;A,B=0,1,2,3,4).
\]

Define the pseudo-scalar
\begin{equation}
\mathbf{i}=F_{0}F_{1}F_{2}F_{3}F_{4}\text{ \quad}\mathbf{i}^{2}=-1\text{
\quad}\mathbf{i}F_{A}=F_{A}\mathbf{i}\text{ \quad}A=0,1,2,3,4 \tag{D6}%
\end{equation}

Define
\begin{equation}
\mathcal{E}_{\mu}=F_{\mu}F_{4} \tag{D7}%
\end{equation}
We can immediately verify that $\mathcal{E}_{\mu}\mathcal{E}_{\upsilon
}+\mathcal{E}_{\upsilon}\mathcal{E}_{\mu}=2\eta_{\mu\upsilon}$. Taking into
account that $\mathbb{R}_{1,3}\simeq\mathbb{R}_{4,1}^{0}$ we can explicitly
exhibit here this isomorphism by considering the map \texttt{j}$:\mathbb{R}%
_{1,3}\rightarrow\mathbb{R}_{4,1}$ generated by the linear extension of the
map \texttt{j}$^{\#}:\mathbb{R}^{1,3}\rightarrow\mathbb{R}_{4,1}$,
$\mathtt{j}^{\#}(F_{\mu})=\mathcal{E}_{\mu}=$ $F_{\mu}F_{4}$, where
$\mathcal{E}_{\mu}$, ($\mu=0,1,2,3$) is an orthogonal basis of $\mathbb{R}%
^{1,3}$. Also \texttt{j}$(1_{\mathbb{R}_{1,3}})=1_{\mathbb{R}_{4,1}^{+}}$,
where $1_{\mathbb{R}_{1,3}}$ and $1_{\mathbb{R}_{4,1}^{+}}$ are the identity
elements in $\mathbb{R}_{1,3}$ and $\mathbb{R}_{4,1}^{+}$. Now consider the
primitive idempotent of $\mathbb{R}_{1,3}\simeq\mathbb{R}_{4,1}^{0}$,
\begin{equation}
\mathrm{e}_{41}=\mathtt{j}(\mathrm{e})=\frac{1}{2}(1+\mathcal{E}_{0}) \tag{D8}%
\end{equation}
and the minimal left ideal $I_{4,1}=\mathbb{R}_{4,1}\mathrm{e}_{41}$.

In what follows we use (when convenient) for minimal idempotents and\ \ the
\ minimal ideals generated by them, the labels involving the notion of
spinorial frames discussed in section 2. Let then, $\Xi_{0}$ be a fiducial
spinorial frame. The elements\footnote{In what follows we use (when
convinient) for minimal idempotents and\ \ the \ minimal ideals generated by
them, the labels involving the notion of spin frames discussed in section 2.}
$Z_{\Xi_{0}}\in I_{4,1}$ can be written analogously to $\Phi\in\mathbb{R}%
_{1,3}\frac{1}{2}(1+E_{0})$ as,
\begin{equation}
Z_{\Xi_{0}}=\sum z_{i}\bar{s}_{i} \tag{D9}%
\end{equation}
where
\begin{equation}
\bar{s}_{1=}\mathrm{e}_{41}\mathrm{,}\text{ }\bar{s}_{2}=\mathcal{E}%
_{1}\mathcal{E}_{3}\mathrm{e}_{41}\mathrm{,}\text{ }\bar{s}_{3}=\mathcal{E}%
_{3}\mathcal{E}_{0}\mathrm{e}_{41}\mathrm{,}\text{ }\bar{s}_{4}=\mathcal{E}%
_{1}\mathcal{E}_{0}\mathrm{e}_{41} \tag{D10}%
\end{equation}
and where
\[
z_{i}=a_{i}+\mathcal{E}_{2}\mathcal{E}_{1}b_{i},
\]
are formally complex numbers, $a_{i},$ $b_{i}\in\mathbb{R}$.

Consider now the element $f_{\Xi_{0}}\in\mathbb{R}_{4,1}$%
\begin{align}
f_{\Xi_{0}}  &  =\mathrm{e}_{41}\frac{1}{2}(1+\mathbf{i}\mathcal{E}%
_{1}\mathcal{E}_{2})\nonumber\\
&  =\frac{1}{2}(1+\mathcal{E}_{0})\frac{1}{2}(1+\mathbf{i}\mathcal{E}%
_{1}\mathcal{E}_{2}), \tag{D11}%
\end{align}
with $\mathbf{i}$ defined as in Eq.(D6).

Since $f_{\Xi_{0}}\mathbb{R}_{4,1}f_{\Xi_{0}}=\mathbb{C}f_{\Xi_{0}}=f_{\Xi
_{0}}\mathbb{C}$ it follows that $f_{\Xi_{0}}$ is a primitive idempotent of
$\mathbb{R}_{4,1}$. We can easily show that each $\Phi_{\Xi_{0}}\in
I_{_{\Xi_{0}}}=\mathbb{R}_{4,1}f_{\Xi_{0}}$ can be written
\[
\Psi_{\Xi_{0}}=\sum_{i}\psi_{i}f_{i},\text{ }\psi_{i}\in\mathbb{C}%
\]%
\begin{equation}
f_{1}=f_{\Xi_{0}},\text{ }f_{2}=-\mathcal{E}_{1}\mathcal{E}_{3}f_{\Xi_{0}%
},\text{ }f_{3}=\mathcal{E}_{3}\mathcal{E}_{0}f_{\Xi_{0}},\text{ }%
f_{4}=\mathcal{E}_{1}\mathcal{E}_{0}f_{\Xi_{0}} \tag{D12}%
\end{equation}
with the methods described in (\cite{18},\cite{19}) we find the following
representation in $\mathbb{C}(4)$ for the generators $\mathcal{E}_{\mu}$ of
$\mathbb{R}_{4,1}\simeq\mathbb{R}_{1,3}$%
\begin{equation}
\mathcal{E}_{0}\mapsto\underline{\gamma}_{0}=\left(
\begin{array}
[c]{cc}%
\mathbf{1}_{2} & 0\\
0 & \mathbf{-1}_{2}%
\end{array}
\right)  \leftrightarrow\mathcal{E}_{i}\mapsto\underline{\gamma}_{i}=\left(
\begin{array}
[c]{cc}%
0 & -\sigma_{i}\\
\sigma_{i} & 0
\end{array}
\right)  \tag{D13}%
\end{equation}
where $\mathbf{1}_{2}$ is the unit $2\times2$ matrix and $\sigma_{i}$,
($i=1,2,3$) are the standard Pauli matrices. We immediately recognize the
$\underline{\gamma}$-matrices in Eq.(D13) as the standard ones appearing,
e.g., in \cite{2}.

The matrix representation of $\Psi_{\Xi_{0}}\in I_{\Xi_{0}}$ will be denoted
by the same letter without the index, i.e., $\Psi_{\Xi_{0}}\mapsto
\mathbf{\Psi}\in\mathbb{C}(4)f$, where
\begin{equation}
f=\frac{1}{2}(1+i\underline{\gamma}_{1}\underline{\gamma}_{2})\text{ \quad
}i=\sqrt{-1}. \tag{D14}%
\end{equation}
We have
\begin{equation}
\mathbf{\Psi}=\left(
\begin{array}
[c]{llll}%
\psi_{1} & 0 & 0 & 0\\
\psi_{2} & 0 & 0 & 0\\
\psi_{3} & 0 & 0 & 0\\
\psi_{4} & 0 & 0 & 0
\end{array}
\right)  ,\text{ }\psi_{i}\in\mathbb{C} \tag{D15}%
\end{equation}
Eqs.(D13, D14, D15) are sufficient to prove that there are bijections between
the elements of the ideals $\mathbb{R}_{1,3}\frac{1}{2}(1+E_{0})$,
$\mathbb{R}_{4,1}\frac{1}{2}(1+\mathcal{E}_{0})$ and $\mathbb{R}_{4,1}\frac
{1}{2}(1+\mathcal{E}_{0})\frac{1}{2}(1+\mathbf{i}\mathcal{E}_{1}%
\mathcal{E}_{2})$.

We can easily find that the following relation exist between $\Psi_{\Xi_{0}%
}\in\mathbb{R}_{4,1}f_{\Xi_{0}}$ and $Z_{\Xi_{0}}\in\mathbb{R}_{4,1}\frac
{1}{2}(1+\mathcal{E}_{0}),$ $\Xi_{0}=(u_{0},\Sigma_{0})$ being a spinorial
frame (see section 1)
\begin{equation}
\Psi_{\Xi_{0}}=Z_{\Xi_{0}}\frac{1}{2}(1+\mathbf{i}\mathcal{E}_{1}%
\mathcal{E}_{2}). \tag{D16}%
\end{equation}

Decomposing $Z_{\Xi_{0}}$ into even and odd parts relatives to the
$\mathbf{Z}_{2}$-graduation of $\mathbb{R}_{4,1}^{0}\simeq\mathbb{R}_{1,3}$,
$Z_{\Xi_{0}}=Z_{\Xi_{0}}^{0}+Z_{\Xi_{0}}^{1}$ we obtain $Z_{\Xi_{0}}%
^{0}=Z_{\Xi_{0}}^{1}\mathcal{E}_{0}$ which clearly shows that all information
of $Z_{\Xi_{0}}$ is contained in $Z_{\Xi_{0}}^{0}$. Then,
\begin{equation}
\Psi_{\Xi_{0}}=Z_{\Xi_{0}}^{0}\frac{1}{2}(1+\mathcal{E}_{0})\frac{1}%
{2}(1+\mathbf{i}\mathcal{E}_{1}\mathcal{E}_{2}). \tag{D17}%
\end{equation}

Now, if we take into account (\cite{49}) that $\mathbb{R}_{4,1}^{0}\frac{1}%
{2}(1+\mathcal{E}_{0})=\mathbb{R}_{4,1}^{00}\frac{1}{2}(1+\mathcal{E}_{0})$
where the symbol $\mathbb{R}_{4,1}^{00}$ means $\mathbb{R}_{4,1}^{00}%
\simeq\mathbb{R}_{1,3}^{0}\simeq\mathbb{R}_{3,0}$ we see that each $Z_{\Xi
_{0}}\in\mathbb{R}_{4,1}\frac{1}{2}(1+\mathcal{E}_{0})$ can be written
\begin{equation}
Z_{\Xi_{0}}=\psi_{\Xi_{0}}\frac{1}{2}(1+\mathcal{E}_{0})\text{ \quad}\psi
_{\Xi_{0}}\in\mathbb{R}_{4,1}^{00}\simeq\mathbb{R}_{1,3}^{0}. \tag{D18}%
\end{equation}
Then putting $Z_{\Xi_{0}}^{0}=\psi_{\Xi_{0}}/2$, Eq.(D18) can be written
\begin{align}
\Psi_{\Xi_{0}}  &  =\psi_{\Xi_{0}}\frac{1}{2}(1+\mathcal{E}_{0})\frac{1}%
{2}(1+\mathbf{i}\mathcal{E}_{1}\mathcal{E}_{2})\nonumber\\
&  =Z_{\Xi_{0}}^{0}\frac{1}{2}(1+\mathbf{i}\mathcal{E}_{1}\mathcal{E}_{2}).
\tag{D19}%
\end{align}

The matrix representation of $Z_{\Xi_{0}}$ and $\psi_{\Xi_{0}}$ in
$\mathbb{C}(4)$ (denoted by the same letter in boldface without index) in the
spin basis given by Eq.(D12) are
\begin{equation}
\mathbf{\Psi}=\left(
\begin{array}
[c]{cccc}%
\psi_{1} & -\psi_{2}^{\ast} & \psi_{3} & \psi_{4}^{\ast}\\
\psi_{2} & \psi_{1}^{\ast} & \psi_{4} & -\psi_{3}^{\ast}\\
\psi_{3} & \psi_{4}^{\ast} & \psi_{1} & -\psi_{2}^{\ast}\\
\psi_{4} & -\psi_{3}^{\ast} & \psi_{2} & \psi_{1}^{\ast}%
\end{array}
\right)  ,\text{ }\mathbf{Z}=\left(
\begin{array}
[c]{cccc}%
\psi_{1} & -\psi_{2}^{\ast} & 0 & 0\\
\psi_{2} & \psi_{1}^{\ast} & 0 & 0\\
\psi_{3} & \psi_{4}^{\ast} & 0 & 0\\
\psi_{4} & -\psi_{3}^{\ast} & 0 & 0
\end{array}
\right)  . \tag{D20}%
\end{equation}

\section{What is a Covariant Dirac Spinor (\emph{CDS})}

As we already know $f_{\Xi_{0}}=\frac{1}{2}(1+\mathcal{E}_{0})\frac{1}%
{2}(1+\mathbf{i}\mathcal{E}_{1}\mathcal{E}_{2})$ (Eq.(D12)) is a primitive
idempotent of $\mathbb{R}_{4,1}\simeq\mathbb{C}(4)$. If $u\in\mathrm{Spin}%
\left(  1,3\right)  \subset\mathrm{Spin}(4,1)$ then all ideals $I_{\Xi_{u}%
}=I_{\Xi_{0}}u^{-1}$ are geometrically equivalent to $I_{\Xi_{0}}$.Now, let
$\mathbf{s(}\Xi_{u})=\{\mathfrak{E}_{0},\mathfrak{E}_{1},\mathfrak{E}%
_{2},\mathfrak{E}_{3}\}$ and $\mathbf{s(}\Xi_{u^{\prime}})=\{\mathfrak{E}%
_{0}^{\prime},\mathfrak{E}_{1}^{\prime},\mathfrak{E}_{2}^{\prime}%
,\mathfrak{E}_{3}^{\prime}\}$ with $\mathbf{s(}\Xi_{u})=u^{-1}\mathbf{s}%
(\Xi_{o})u$, $\mathbf{s(}\Xi_{u^{\prime}})=u^{\prime-1}\mathbf{s}(\Xi
_{o})u^{\prime}$ be \ two arbitrary basis for $\mathbb{R}^{1,3}\subset
\mathbb{R}_{4,1}$. From Eq.(D13) we can write
\begin{equation}
I_{\Xi_{u}}\ni\Psi_{\Xi_{u}}=\sum\psi_{i}f_{i},\text{ and }I_{\Xi_{u}^{\prime
}}\ni\Psi_{\Xi_{u^{\prime}}}=\sum\psi_{i}^{\prime}f_{i}^{\prime}, \tag{E1}%
\end{equation}
where
\[
f_{1}=f_{\Xi_{u}},\text{ \quad}f_{2}=-\mathfrak{E}_{1}\mathfrak{E}_{3}%
f_{\Xi_{u}},\text{ \quad}f_{3}=\mathfrak{E}_{3}\mathfrak{E}_{0}f_{\Xi_{u}%
},\text{ \quad}f_{4}=\mathfrak{E}_{1}\mathfrak{E}_{0}f_{\Xi_{u}}%
\]
and
\[
f_{1}^{\prime}=f_{\Xi_{u^{\prime}}},\text{ \quad}f_{2}^{\prime}=-\mathfrak{E}%
_{1}^{\prime}\mathfrak{E}_{3}^{\prime}f_{\Xi_{u^{\prime}}},\text{ \quad}%
f_{3}^{\prime}=\mathfrak{E}_{3}^{\prime}\mathfrak{E}_{0}^{\prime}%
f_{\Xi_{u^{\prime}}},\text{ \quad}f_{4}=\mathfrak{E}_{1}^{\prime}%
\mathfrak{E}_{0}^{\prime}f_{\Xi_{u^{\prime}}}%
\]

Since $\Psi_{\Xi_{u^{\prime}}}=\Psi_{\Xi_{u}}(u^{\prime-1}u)^{-1}$, we get
\[
\Psi_{\Xi_{u}^{\prime}}=\sum_{i}\psi_{i}(u^{\prime-1}u)^{-1}f_{i}^{\prime
}=\sum_{i,k}S_{ik}[(u^{-1}u^{\prime})]\psi_{i}f_{k}=\sum_{k}\psi_{k}^{\prime
}f_{k}.
\]

Then
\begin{equation}
\psi_{k}^{\prime}=\sum_{i}S_{ik}(u^{-1}u^{\prime})\psi_{i}, \tag{E2}%
\end{equation}
where $S_{ik}(u^{-1}u^{\prime})$ are the matrix components of the
representation in $\mathbb{C}(4)$ of ($u^{-1}u^{\prime}$)$\in\mathrm{Spin}%
_{1,3}^{e}$. As proved in (\cite{18},\cite{19}) the matrices $S(u)$ correspond
to the representation $D^{(1/2,0)}\oplus D^{(0,1/2)}$ of $SL(2,\mathbb{C}%
)\simeq\mathrm{Spin}_{1,3}^{e}$.

We remark that all the elements of the set \{$I_{\Xi_{u}}$\} of the ideals
geometrically equivalent to $I_{\Xi_{0}}$ under the action of $u\in
\mathrm{Spin}_{1,3}^{e}\subset\mathrm{Spin}_{4,1}^{e}$ have the same image
$I=\mathbb{C}(4)f$ where $f$ is given by Eq.(D11), i.e.,
\begin{equation}
f=\frac{1}{2}(1+\underline{\gamma}_{0})(1+i\underline{\gamma}_{1}%
\underline{\gamma}_{2}),\text{ \quad}i=\sqrt{-1}, \tag{E3}%
\end{equation}
where $\underline{\gamma}_{\mu},$ $\mu=0,1,2,3$ are the Dirac matrices given
by Eq..(D14).

Then, if
\begin{align}
\gamma &  :\mathbb{R}_{4,1}\rightarrow\mathbb{C}(4)\equiv\mathrm{End}%
(\mathbb{C}(4)f),\nonumber\\
x  &  \mapsto\gamma(x):\mathbb{C}(4)f\rightarrow\mathbb{C}(4)f \tag{E4}%
\end{align}
it follows that
\begin{equation}
\gamma(\mathfrak{E}_{\mu})=\gamma(\mathfrak{E}_{\mu}^{\prime})=\underline
{\gamma}_{\mu},\text{ }\gamma(f_{\mu})=\gamma(f_{\mu}^{\prime}) \tag{E5}%
\end{equation}
for all $\{\mathfrak{E}_{\mu}\}$, $\{\mathfrak{E}_{\mu}^{\prime}\}$ such that
$\mathfrak{E}_{\mu}^{\prime}=(u^{\prime-1}u)\mathfrak{E}_{\mu}(u^{\prime
-1}u)^{-1}$. Observe that \textit{all information} concerning the geometrical
images of the spinorial frames $\Xi_{u}$, $\Xi_{u^{\prime}},...,$ under
$\mathbf{L}^{\prime}$disappear in the matrix representation of the ideals
$I_{\Xi_{u}},$ $I_{\Xi_{u^{\prime}}},....,$ in $\mathbb{C(}4\mathbb{)}$ since
all these ideals are mapped in the same ideal $I=\mathbb{C(}4\mathbb{)}f$.

With the above remark and taking into account the definition of algebraic
spinors given in section 2.3 and Eq.(E2) we are lead to the following

\begin{definition}
A covariant Dirac spinor (CDS) for $\mathbb{R}^{1,3}$ is an equivalence class
of pairs $(\Xi_{u}^{m},\mathbf{\Psi})$, where $\Xi_{u}^{m}$ is a matrix
spinorial frame associated to the spinorial frame $\Xi_{u}$ through the
$S(u^{-1})\in D^{(\frac{1}{2},0)}\oplus D^{(0,\frac{1}{2})}$ representation of
$\mathrm{Spin}_{1,3}^{e},$ $u\in\mathrm{Spin}_{1,3}^{e}$.We say that
$\mathbf{\Psi,\Psi}^{\prime}\in\mathbb{C}(4)f$ are equivalent and write
\end{definition}

\begin{equation}
(\Xi_{u}^{m},\mathbf{\Psi})\sim(\Xi_{u^{\prime}}^{m},\mathbf{\Psi}^{\prime})
\tag{E6}%
\end{equation}
\emph{if and only if }%

\begin{equation}
\mathbf{\Psi}^{\prime}=S(u^{\prime-1}u)\mathbf{\Psi,}\text{ }u\mathbf{s(}%
\mathfrak{\Xi}_{u})u^{-1}=u^{\prime}\mathbf{s}(\Xi_{u^{\prime}})u^{\prime
^{-1}} \tag{E7}%
\end{equation}

\begin{remark}
The definition of CDS just given agrees with that given in \cite{11} except
for the irrelevant fact that there, as well as in the majority of Physics
textbook's, authors use as the space of representatives of a CDS a complex
four-dimensional space $\mathbb{C}^{4}$ instead of $I=\mathbb{C}(4)f$.
\end{remark}

\end{document}